\documentclass{sig-alternate}
\usepackage{cite}
\usepackage[]{algorithm2e}
\usepackage{subfig}
\usepackage{graphicx}
\usepackage{epstopdf}
\usepackage{color}
\usepackage{multirow}
\newcommand\independent{\protect\mathpalette{\protect\independenT}{\perp}}
\def\independenT#1#2{\mathrel{\rlap{$#1#2$}\mkern2mu{#1#2}}}

\begin{document}
\title{Non-parametric Causality Detection: An Application to Social Media and Financial Data}
\numberofauthors{3} %  in this sample file, there are a *total*
\author{
% 1st. author
\alignauthor
Fani Tsapeli \\
       \affaddr{School of Computer Science}\\
       \affaddr{University of Birmingham}\\
       \affaddr{Edgbaston B15 2TT Birmingham, UK}\\
       \email{t.tsapeli@cs.bham.ac.uk}
% 2nd. author
\alignauthor
Mirco Musolesi \\
       \affaddr{Department of Geography}\\
       \affaddr{University College London}\\
       \affaddr{Gower Street WC1E 6BT London, UK}\\
       \email{m.musolesi@ucl.ac.uk}
% 3rd. author
\alignauthor
Peter Tino \\
       \affaddr{School of Computer Science}\\
      \affaddr{University of Birmingham}\\
      \affaddr{Edgbaston B15 2TT Birmingham, UK}\\
      \email{p.tino@cs.bham.ac.uk}
}
\maketitle
\begin{abstract}
According to behavioral finance, stock market returns are influenced by emotional, social and psychological factors. Several recent works support this theory by providing evidence of correlation between stock market prices and collective sentiment indexes measured using social media data. However, a pure correlation analysis is not sufficient to prove that stock market returns are influenced by such emotional factors since both stock market prices and collective sentiment may be driven by a third unmeasured factor. Controlling for factors that could influence the study by applying multivariate regression models is challenging given the complexity of stock market data. False assumptions about the linearity or non-linearity of the model and inaccuracies on model specification may result in misleading conclusions.

In this work, we propose a novel framework for causal inference that does not require any assumption about a particular parametric form of the model expressing statistical  relationships among the variables of the study and can effectively control a large number of observed factors. We apply our method in order to estimate the causal impact that information posted in social media may have on stock market returns of four big companies. Our results indicate that social media data not only correlate with stock market returns but also influence them.

\end{abstract}



 % NOT required for Proceedings

\section{Introduction}

We are living in the era of social media, 
using tools such as Facebook, Twitter and blogs to communicate with our friends, to share our experiences and to express our opinion and emotions. Recently, mining and analyzing this kind of data has emerged as an area of great interest for both the industrial and academic communities. Several studies have examined the ability of social media to serve as crowd-sensing platforms. For example, authors in~\cite{asur2010predicting} demonstrate that social media can monitor the popularity of products or services and predict their future revenues. Evidence has been found that social media can be used to predict election results \cite{tumasjan2010predicting} or even stock market prices \cite{bollen2011twitter}.

Most studies so far have focused on using social media data as early indicators of real-world events. But to what extent do opinions expressed through social media actually have a \textit{causal} influence on the examined events? For example, are stock market prices influenced by the opinions and sentiments that are reported in social media, or is it the case that stock market prices and sentiments are driven only by other (e.g. financial) factors? Would the results have been different if we could manipulate social media data? In order to answer such questions a causality study is required. 

Some recent studies have examined the ability of social media to influence real-world events by applying randomized control trials. For example, authors in~\cite{Bond2012} examine the effect of political mobilization messages by using Facebook to deliver such messages to a randomly selected population; the effect of the messages is measured by comparing the real-world voting activity of this group with the voting activity of a control group. Similarly, in \cite{Muchnik2013} authors use randomized trials in order to examine the social influence of aggregated opinions posted in a social news website. Randomized control trials are a reliable technique for conducting causal inference studies \cite{concato2000randomized}. However, their applicability is limited since they require scientists to gather data using experimental procedures and do not allow the exploitation of the large amount of observational data. In many cases, it is not feasible to apply experimental designs or it is considered unethical. 

In this work, we study the causal impact of social, psychological and emotional factors on stock market prices of big companies using observational data collected through Twitter. Twitter enables us to capture people sentiments and opinions about traded assets and their reactions on related news and events. Previous works have demonstrated that social media data correlate with stock market prices~\cite{preis2010complex, bollen2011twitter, zhang2011predicting, zhang2012predicting}. These studies were predominantly based on correlation or Granger causality analysis. Granger causality tests the ability of a time-series to predict values of another one \cite{granger1988some}. However, it cannot be used to discover real causality. A positive result on a Granger causality test does not necessarily imply that there is a causal link between the examined time-series since both the examined time-series may be influenced by a third variable (\textit{confounding bias}). Multivariate regression techniques can be applied in order to control for confounding bias. Some studies attempt to improve the accuracy of stock market prediction models by applying multivariate regression \cite{mao2011predicting}. However, the focus of these works is on prediction rather than on causal inference. Applying regression models for causal inference suffers from two main limitations. First, stock market prices can be influenced by a large number of factors such as stock market prices of other companies \cite{kenett2010dominating, chi2010network}, foreign currency exchange rates and commodity prices. Such factors may also influence people sentiments. Consequently, to eliminate any confounding bias one is required to include a large number of predictors in the regression model. Estimation of regression coefficients in a model with a large number of predictors can be challenging.  When data dimensionality is comparable to the sample size, noise may dominate the 'true' signal, rendering the study infeasible~\cite{fan2014challenges}. Second, inaccuracies in model specification, estimation or selection may result in invalid causal conclusions.

Given the limitations of parametric methods, we propose a novel framework for causal inference in time-series that is based on \textit{matching design}~\cite{william2002experimental, MatchingSurvey}. This technique attempts to eliminate confounding bias by creating pairs of \textit{similar} treated and untreated objects, i.e. objects with similar values on baseline characteristics that could influence the causality study. Thus, the effect of an event is estimated by comparing each object exposed to an event with a \textit{similar} object that has not been exposed. Matching design bypasses the limitations of regression-based methods since it does not require specification of a model class. However, it cannot be applied in time-series since it assumes that the objects of the study are realizations of i.i.d variables. We reformulate the concept of matching design to make it suitable for causal inference on time-series data.  In our case the time-series collection includes  \textit{treatment} time-series $X$, \textit{response} time-series $Y$ and a set of time-series $\mathbf{Z}$ which contain characteristics relevant to the study. The units of our study correspond to time-samples; the $t^{th}$ unit is characterized by a treatment value $X(t)$, a response value $Y(t)$ and a set of values representing baseline characteristics $\mathbf{Z(t)}$. 
We assess the causal impact of a time-series $X$ on $Y$ by comparing different units (i.e. time-samples) on $Y$ after controlling for characteristics captured in $\mathbf{Z}$. As explained in Section \ref{sec:mechanism}, our methodology assures that the objects are uncorrelated, which is a weaker version of the independence assumption requirement of the matching design. We apply our framework in order to estimate the causal impact that the sentiment of information posted in social media may have on traded assets. In detail, we estimate a daily sentiment index (\textit{treatment} time-series) based on information posted in Twitter and we assess its impact on daily stock market closing prices (\textit{response} time-series) of four big technological companies after controlling for other factors (set of time-series $\mathbf{Z}$) that may influence the study, such as the performance of other big companies.

In summary, the contribution of this work is twofold:

\begin{enumerate}
\item We propose a causal inference framework for time-series that can be applied to high-dimensional data without imposing any restriction on the model class describing the associations among the data. We demonstrate, using synthetic data, that our methodology is more effective on detecting true causality compared to other methods that have been applied so far, for causal inference in time-series. 

\item We apply our method in order to quantify the causal impact of emotional and psychological factors, captured by social media, on stock market prices of four technological companies. To the best of our knowledge, this is the first study that measures the causal influence of such factors on finance. It should be noted that,  since all the examined companies belong to the technological sector, our findings cannot be directly generalized for any company. 
\end{enumerate}

The rest of this paper is organized as follows. In Section \ref{sec:background} we discuss the main methodologies that are used for causal inference. In Section \ref{sec:mechanism} we present the proposed framework. In Section \ref{sec:evaluation} we evaluate our approach on synthetic data, in conjunction with other methods that have been previously applied for causal inference in time-series. Moreover, we apply our method in order to assess the causal impact of information posted in Twitter on stock market prices of specific companies. In Section \ref{sec:related-work} we discuss some relevant works which attempt to uncover the relationship between social media data and stock market movement. In Section \ref{sec:discussion} we summarize the advantages and limitations of the proposed approach and we discuss its computational cost. Finally, Section \ref{sec:conclusions} concludes the paper by summarizing our contributions.

\section{Background on Causal Inference}
\label{sec:background}

Causal analysis attempts to understand whether differences on a specific characteristic $Y$ within a population of \textit{units} are caused by a factor $X$. $Y$ is called \textit{response}, \textit{effect} or \textit{outcome} variable and $X$ \textit{treatment} variable or \textit{cause}.  \textit{Units} are the basic objects of the study and they may correspond to humans, animals or any kind of experimental objects. 

\subsection{Potential Outcome Framework}

The key concept on causation theory is that given a unit $u$, the value of the corresponding response variable  $Y(u)$ can be manipulated by changing the value of the treatment variable $X(u)$~\cite{Holland86, morgan2014counterfactuals}.  In this paper, we will consider $X(u)$ as a binary treatment variable. Hence, there will be two treatments: $x_1$ for treated units and $x_0$ for untreated. We denote by $Y_1(u)$ the value of $Y$ when $X(u)=x_1$ and $Y_0(u)$ when $X(u)=x_0$. In order to test the effect of $x_1$ on unit $u$, we need to estimate the quantity $Y_1(u) - Y_0(u)$. The fundamental problem of causal inference is that \textit{it is impossible to observe both $Y_1(u)$ and $Y_0(u)$ on the same unit $u$ and, therefore, it is impossible to measure the real causal effect of $x_1$ on the unit.}~\cite{Holland86, morgan2014counterfactuals}. Thus, the average effect of a treatment $X$ is estimated by comparing a population of objects that received the treatment $x_1$ with a population that received the treatment $x_0$ and evaluating the corresponding average values of the effect variable $Y$. We denote by  $Y_1$ and $Y_0$ random variables representing the outcome variable when the treatments $x_1$ and $x_0$ are applied, respectively. We also define a (random) variable $DY = Y_1 - Y_0$. Then, the average treatment effect (ATE) of $x_1$  is estimated as the expected value $E\{DY\}$. 
 
The average treatment effect can be estimated only if the following three assumptions are satisfied:

\begin{enumerate}
\item the effect differences are i.i.d. realizations of  $DY$.
\item the observed outcome in one unit is independent from the treatment received by any other unit (\textit{Stable Unit Treatment Value Assumption - SUTVA}).
\item the assignment of units to treatments is independent from the outcome (\textit{ignorability}). Ignorability can be formally expressed as $Y_1 \independent X$, $Y_0 \independent X$. The assumption of ignorability requires that all the units have equal probability to be assigned to a treatment. If this assumption does not hold, the units that received a treatment $x_1$ may systematically differ from units that did not receive such a treatment. In such a case  the average value of the outcome variable of the treated units  could be different from that of other units, even if the treatment had not been received at all. 
\end{enumerate}

In experimental studies, ignorability can be achieved by randomly assigning units to treatments. However, in observational studies this is not feasible. Instead, the average treatment effect can be estimated by relaxing ignorability to \textit{conditional ignorability}. According to conditional ignorability assumption, the treatment assignment is independent from the outcome, conditional on a set of confounding  variables $\mathbf{H}$. Variables $\mathbf{H}$ represent baseline characteristics of the units that are considered relevant for the study (e.g. in a medical study that examines the impact of a drug, baseline characteristics could be the previous health condition of the units (in this case patients), their age etc.). Thus, conditional ignorability is expressed as $Y_1 \independent X|\mathbf{H}$, $Y_0 \independent X|\mathbf{H}$. The variables that must be included in the set $\mathbf{H}$ in order to achieve conditional ignorability are also called \textit{confounding variables}. Conditional ignorability cannot be achieved when one or more confounding variables are unobserved. The main limitation of all non-experimental causality studies is that the possibility that important confounding variables are missing cannot be eliminated. Latent variable models \cite{bentler1980multivariate} and analysis on the sensitivity of the conclusions on missing confounding variables \cite{rosenbaum1983assessing} have been proposed to handle this issue. 

There are two main methodologies that are applied in order to achieve conditional ignorability: \textit{regression} and \textit{matching}~\cite{william2002experimental}. Regression expresses the outcome variable $Y$ as a function of the treatment variable $X$ and the set of variables $\mathbf{H}$~\cite{freedman1997association, imbens2004nonparametric}. Linear models are usually applied. Methods based on regression require  scientists to correctly specify a regression model. These methods are affected by the typical problems of parametric approaches to causality detection (i.e. model misspecification, overfitting and poor performance on high-dimensional datasets).

Matching comprises a more flexible methodology for causal inference in observational data since it does not require the specification of a model class~\cite{MatchingSurvey}. Conditional ignorability is achieved by creating sub-population within which the values of the confounding variables $\mathbf{H}$ are the same or similar. Thus, considering a set $G$ of pairs of treated and untreated (\textit{control}) units $(u, v)$ such that $\mathbf{H}(u)  \approx \mathbf{H}(v)$, we can estimate the average treatment effect (ATE) as 
\begin{equation}
\widehat{E}\{DY\} = \frac{\sum_{(u, v) \in G} Y(u) - Y(v)}{|G|},
\label{eq:ATE}
\end{equation}
where $|G|$ denotes the size of $G$.
Scientists need to assess the degree of similarity between the matched treated and control units. Similarity relation ($\approx$) can be assessed by estimating the standardized mean difference for each confounding variable between matched treated and control units, or by applying graphical methods such as quantile-quantile plots, cumulative distribution functions plots, etc. \cite{austin2009relative, austin2008goodness, austin2008assessing, harder2010propensity}. If sufficient balance has not been achieved, the applied matching method needs to be revised.

\subsection{Directed Acyclic Graphs}

Pearl \cite{pearl2010introduction, pearl2000causality, pearl1995causal} proposed the use of directed acyclic graphs (DAGs) for representing causal relationships. In causal graphs, nodes represent the variables of the experiment. If $\mathbf{P}$ a set of the direct predecessors (\textit{parents}) of a node $Y$, a direct arrow from a node $Q$ (can represent $X$ or one of the variables of set $\mathbf{Z}$) to $Y$ will exist only if $Y \not\independent Q | \mathbf{P} \setminus \{Q\}$. A direct arrow from a node $Q$ to a node $Y$ represents a causal relationship between the two variables (i.e. $Q$ causes $Y$). 

Pearl also introduces a graphical criterion for defining a \textit{sufficient set} of variables that need to be controlled in order to achieve conditional ignorability when testing the causal impact of a variable $X$ on a variable $Y$ (\textit{back-door criterion}). According to this rule, a subset $\mathbf{H}$ of  variables  is \textit{sufficient} if no element of $\mathbf{H}$ is a descendant of $X$ and the elements of $\mathbf{H}$ block all paths from $X$ to $Y$ that end with an arrow to $X$ (\textit{back-door paths}). The intuition behind this criterion is that back-door paths from $X$ to $Y$ represent spurious associations and therefore need to be excluded in order to obtain unbiased estimation of the causal effect of $X$ on $Y$. 

Causal graphs differ from Bayesian graphs since, in a causal graph, an arrow from $Q$ to $Y$ denotes that $Y$ values would change if we could manipulate the values of $Q$. However, this hypothesis cannot be tested in observational studies. Instead, a graph is created either by utilizing prior knowledge about the structure of the model or by conducting conditional independence tests on the observational data \cite{kalisch2007estimating, li2009controlling, spirtes2001anytime, colombo2012learning}. The correctness of the graph can be assessed by fitting the observational data to a system of structural equations derived from the graph (i.e. each variable is regressed on all its direct predecessors) \cite{pearl1995causal}. 

\subsection{Causality on Time-series}

Causality studies on time-series have been largely based on Granger causality \cite{granger1969investigating}. The Granger causality test examines if past values of one variable are useful in prediction of future values of another variable. In detail, a time-series $X$ Granger causes a time-series $Y$ if modeling $Y$ by regressing it on past values of both $Y$ and $X$ results in reduced residual noise compared to a simple autoregressive model. Granger causality does not test real causality since the conditional ignorability assumption is not satisfied, i.e., the values of both treatment variable $X$ and control variable $Y$ may be driven by a third variable. Moreover, it considers only linear relationships. Granger causality has been extended to handle multivariate cases \cite{barrett2010multivariate} as well as non-linear cases \cite{zhang2012kernel, nawrath2010distinguishing}. However, as it was mentioned also at the introduction, inaccuracies on model specification may result in misleading conclusions. In \cite{peters2013causal} the authors propose an additional model check procedure after fitting a model in order to reduce the amount of false positive causality results. Moreover, in \cite{entner2010causal} authors propose a time-series causality framework based on graph models. The main advantage of the proposed method is the ability to model latent variables (i.e. unobserved confounding variables). However, this method performs worse than Granger causality for large time-series sample sizes. 

Non-parametric approaches (i.e. approaches that do not require the specification of a model class) for causal inference in time-series have also been proposed. Schreiber introduced \textit{transfer entropy} \cite{schreiber2000measuring} in order to examine whether, given a set of multivariate time-series $\mathbf{S}$, the uncertainty about a time-series $Y \in \mathbf{S}$ is reduced by learning the past of a time-series $X \in \mathbf{S}$, when the past of the other time-series in $\mathbf{S}$ is known. Transfer entropy is a model-free equivalent of Granger causality \cite{barnett2009granger}. The main limitation of this approach is that it requires the estimation of a large number of conditional probability densities which is particularly challenging on continuous datasets \cite{zhang2012kernel}. Runge et al. propose the combination of transfer entropy with directed acyclic graphs in order to reduce the number of densities that need to be estimated \cite{pompe2011momentary, runge2012escaping}. In detail, causality is estimated by examining whether uncertainty about time-series $Y$ can be reduced by learning the past of $X$, when the \emph{parents} of $Y$ and $X$ are known. The parents $P_Y$ of a time-series $Y$ are defined as the minimum set of graph-nodes which separate $Y$ with the past of $\mathbf{S} \setminus \{P_Y\}$. Although this modification reduces significantly the number of density estimations that are required, the dimensionality of the dataset may still be high (i.e. the number of parents of $Y$ and $X$ may be very large) which imposes challenges on the estimation of transfer entropy.

\section{Proposed Mechanism}
\label{sec:mechanism}

\begin{figure}
  \centering
\includegraphics[width=0.45\textwidth]{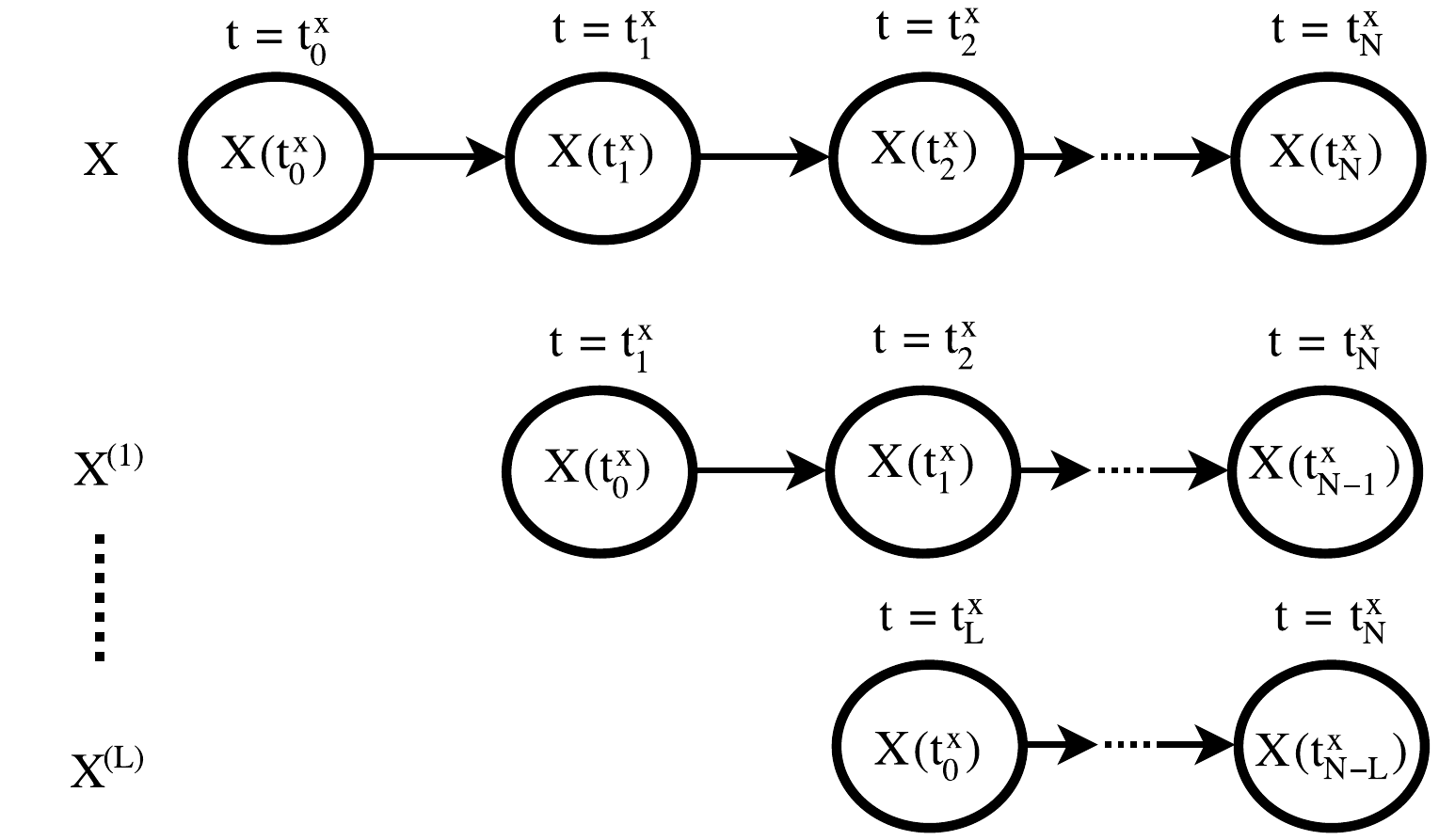}
  
\caption{Graphical representation of time-series $X$ along with its first $L$ lagged versions.}
\label{fig:timeSeriesExample}
\end{figure}

Given the previously discussed limitations on existing methodologies for causality discovery in time-series, we propose a novel framework that enables causal inference in time-series  data that is based on matching design and therefore does not require specification of a model class. The proposed framework also requires only few conditional independence tests, thus it can handle more effectively high-dimensional data. Denote by $Y=\{Y(t^y_i):i=0, 1..., N\}$ and $X=\{X(t^x_i):i=0, 1..., N\}$ the time-series that represent the effect and the cause, respectively and by $\mathbf{Z}=\{\mathbf{Z}(t^z_i):i=0, 1..., N\}$ a set of time-series representing other characteristics relevant for the study. In this study, we consider $X$ as a binary treatment variable. Matching design has been proposed also for non-binary treatments \cite{lu2001matching, hirano2004propensity}, however, extending our framework to non-binary cases is out of the scope of this study. Let us also denote by $Y^{(l)}$, $X^{(l)}$ and $\mathbf{Z}^{(l)}$ the $l$-lagged versions of the time series $Y$, $X$ and $\mathbf{Z}$, respectively (i.e., if $X(t^x_i)$ the $i$-th sample of $X$, $X^{(l)}(t^x_i) = X(t^x_{i-l})$). At Figure \ref{fig:timeSeriesExample} we provide a graphical representation of time-series $X$, $X^{(1)}$, ..., $X^{(L)}$. We define a maximum lag value $L$ and a set of time-series $\mathbf{S}$ =\{$Y$, $Y^{(1)}$, ..., $Y^{(L)}$, $X$, $X^{(1)}$, ..., $X^{(L)}$, $\mathbf{Z}$, $\mathbf{Z}^{(1)}$, ..., $\mathbf{Z}^{(L)}$\}. 

\begin{figure}
  \centering
\includegraphics[width=0.4\textwidth]{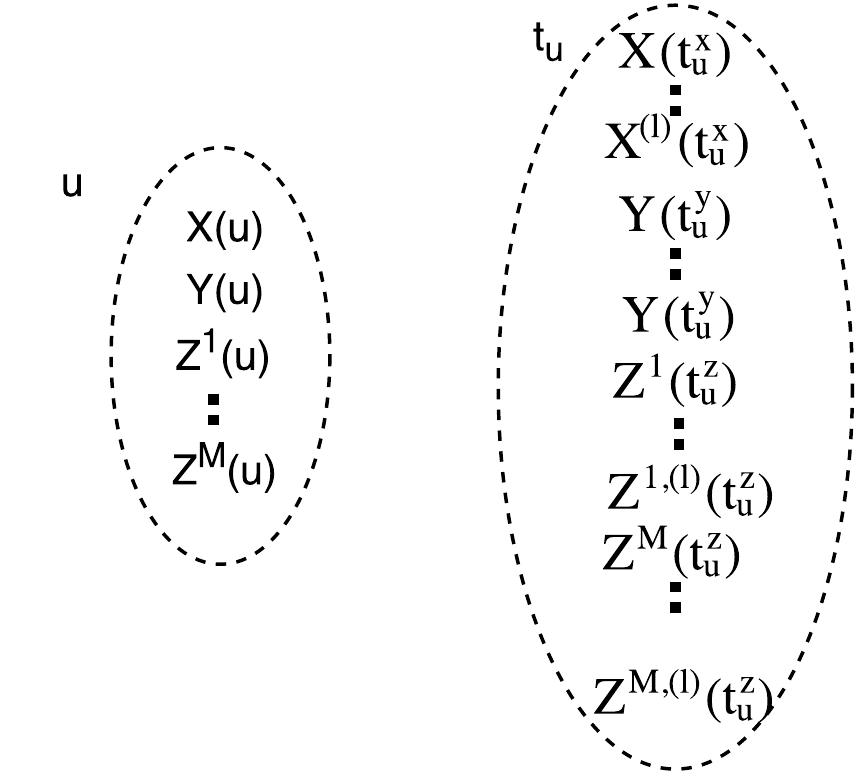}
  
\caption{Graphical representation of units. On the left side, $u$ represents a unit on a traditional causality study, characterized by its treatment value $X(u)$, its response value $Y(u)$ and $M$ other characteristics $Z^1(u)$, $Z^2(u)$..., $Z^M(u)$. On the right side, $t_u$ represents a unit on our time-series matching design framework. The unit is characterized by the time-series values of set $\mathbf{S}$ at $u-th$ time-sample.}
\label{fig:unitsExample}
\end{figure}

As we previously discussed at Section \ref{sec:background}, the units of a study traditionally correspond to experimental objects and the variables of the study describe the characteristics of the units as well as the treatment they have received and the corresponding outcome. In our framework, the units of the study correspond to time-samples of the set of times-series $\mathbf{S}$. For example, consider a study that examines the effects of an industry on the pollution level in a region based on weekly measurements. In this case, a unit of the study corresponds to one week and the variables of the study to weekly pollution measurements, industrial wastes and other relevant characteristics. In Figure \ref{fig:unitsExample} we graphically depict the notion of a unit in our time-series matching design framework in comparison with the traditional notion of unit on causality studies. In the rest of this paper, the terms 'unit' and 'time-sample' will be used interchangeably. 

In order to build a graph, we examine the dependencies between the variables $X$, $Y$ and all the other variables of the set $\mathbf{S}$. In order to examine if two time-series $X$ and $Y$ are independent (assuming that the time-series are stationary in the first two moments) we can estimate the Pearson correlation coefficient as follows: 

\begin{equation*}
	r_{xy} = \frac{\sum_{u=0}^N (X(t_u^x) - \bar{X})(Y(t_u^y) - \bar{Y})}{\sqrt{\sum_{u=0}^N (X(t_u^x) - \bar{X})^2 \sum_{u=0}^N (Y(t_u^y) - \bar{Y})^2}}
\end{equation*}
with $\bar{X}$, $\bar{Y}$ the sample means of $X$, $Y$ respectively. Vanishing correlation could be considered as indication of independence between the examined time-series. Alternatively, Spearman rank correlation or mutual information could be used in order to examine the dependencies between time-series.  

In a directed acyclic graph representing a Bayesian network, a arrow from a variable $W$ to a variable $Q$ is added only if $Q$ is dependent of $W$, conditional on all direct predecessors of $Q$. In our graph representation, we relax this condition as follows:

A arrow from a lagged node $W^{(l)}$ (including lag 0) to a non-lagged node $Q$ exists if:

\begin{itemize}
\item $W^{(l)}$ precedes temporally $Q$, i.e., $t_u^w<t_u^q$, for any $u$; and
\item $Q \not\independent W^{(l)} | \mathbf{P}^m\cap (W, W^{(1)}, ..., W^{(m)})$, where $\mathbf{P^m}$ is the set of the direct predecessors of $Q$ with maximum lag $m$ and $m<l$.
\end{itemize} 

Thus, in our graph representation, a direct edge between two nodes indicates a dependence but not a necessarily a causal link. Causality will be examined by applying the matching design framework, where the direct predecessors of the treatment and outcome time-series will serve as the confounding variables of the study. The main advantage of the proposed framework is the requirement of a significantly lower number of densities estimations and conditional independence testing compared to other causal inference approaches on time-series \cite{schreiber2000measuring, pompe2011momentary, runge2012escaping}. In what follows we will discuss the details of our methodology and how the three general assumptions of causality studies (discussed in Section \ref{sec:background}) are addressed.

\begin{algorithm}[htp]
 \KwData{The set of time-series $\mathbf{S}$}
 \KwResult{ The set of confounding variables $\mathbf{H}$ }
\SetKwFunction{algo}{algo}\SetKwFunction{predecessors}{predecessors}
{\color{blue}\tcc{Find the parents of Y.}}
$\mathbf{P1}:=\predecessors{$\mathbf{S}$, $Y$}$\;
{\color{blue}\tcc{Find the parents of X.}}
$\mathbf{P2}:=\predecessors{$\mathbf{S}$, $X$}$\;
{\color{blue}\tcc{Find the common parents of X, Y.}}
$\mathbf{H}:= \mathbf{P1}\cap \mathbf{P2}$\;

  \SetKwFunction{myproc}{Procedure}{}{}

{\color{blue}\tcc{This procedure returns a set $\mathbf{P}$ of the direct predecessors of node $Q$. $\mathbf{P}$ is a subset of $\mathbf{S}$.}}
  \myproc{\predecessors{$\mathbf{S}$, $Q$}}{\\
$\mathbf{P}:=\{\}$\;
\For{i=0 to L}{
{\color{blue}\tcc{For all zero-lagged time-series}}
  \For{all $S^{(0)} \in \mathbf{S}$}{
  {\color{blue}\tcc{Find the lagged versions of $S$ which are also parents of $Q$.}}
	 $\mathbf{B}:=(S^{(0)}, ..., S^{(i-1)}) \cap \mathbf{P}$\;
	 \If{({$Q \not\independent S^{(i)}|\mathbf{B}$} \textbf{and} {$S^{(i)}$ precedes $Q$})}{
        $\mathbf{P}:=\mathbf{P} \cup S^{(i)} $\;
   }
  }
 }
return $\mathbf{P}$\;}
 \caption{Defining the set of confounding variables.}
\label{Alg:predecessors}
\end{algorithm}

\begin{figure}
  \centering
\includegraphics[width=0.4\textwidth]{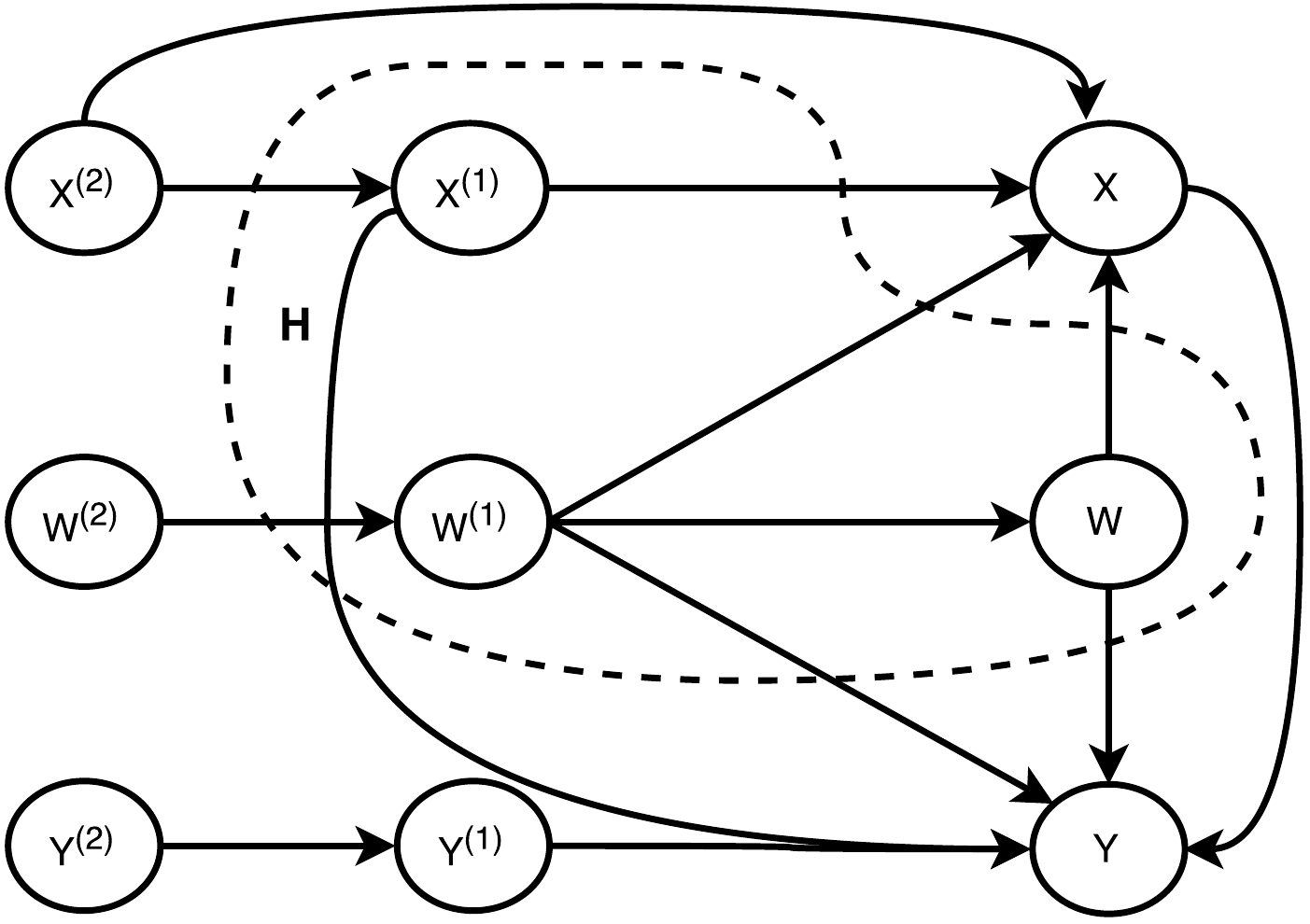}
  
\caption{Example graph depicting the resulted set $\mathbf{H}$ when the impact of $X$ on $Y$ is examined. At this example, X precedes temporally Y and W precedes X. The maximum examined time-lag $L$ is 2. }
\label{fig:exampleGraphHset}
\end{figure}

\textit{Conditional Ignorability Assumption:} We apply the Algorithm \ref{Alg:predecessors} in order to find the set of time-series $\mathbf{H}$ that need to be controlled in order to satisfy the conditional ignorability assumption. According to our method, the resulted set contains all the direct predecessors that nodes $X$ and $Y$ have in common. In figure \ref{fig:exampleGraphHset}, we depict the resulted set $\mathbf{H}$ of an example graph comprised by time-series $X$, $Y$ and $W$ as well as its lagged versions, with maximum lag $L=2$. The parents of $X$ and $Y$ are selected by conducting conditional independence tests as described in Algorithm \ref{Alg:predecessors}. For example, the arrow from $X^{(1)}$ to $X$ denotes that $X \not\independent X^{(1)}$ and the arrow from $X^{(2)}$ to $X$ that $X \not\independent X^{(2)} | X^{(1)}$. Similarly, the arrow from $W$ to $X$ denotes that $X \not\independent W$ and the arrow from $W^{(1)}$ to $X$ denotes that $X \not\independent W^{(1)} | W$. The lack of arrow from $W^{(2)}$ to $X$ denotes that $X \independent W^{(2)} | W, W^{(1)}$. $\mathbf{H}$ includes all the common parents of $X$ and $Y$. Thus, all the variables that are correlated both with $X$ and $Y$ time-series are included; hence, the set $\mathbf{H}$ is sufficient. However $\mathbf{H}$ may include also redundant time-series, i.e., some of the time-series included at $\mathbf{H}$ may not correlate with $X$ or $Y$ conditional to a subset of $\mathbf{H}$. In causality studies based on regression, including redundant predictors on the model could result in overfitting and would jeopardize the validity of the conclusions. Moreover, the application of methods based on conditional independence tests using information theoretic approaches would be challenged by the inclusion of redundant covariates since it would require conditioning on large sets of variables. In contrast, studies based on matching are less affected by the inclusion of redundant confounding variables (spurious correlations). Several methods that enable matching on a large number of confounding variables have been proposed \cite{diamond2013genetic, sekhon2009opiates, rubin1997estimating}. In addition, scientists are able to apply balance diagnostic tests in order to assess if any confounding bias has been adequately eliminated; consequently, false conclusions due to confounding bias can be diminished. Following the matching design, the set of time-series $\mathbf{H}$ is controlled by creating a set of pairs of time-samples $G$ where each $u$-th time-sample with a positive treatment value $X(t^x_u)$ is matched with a $v^{th}$ time-sample with zero treatment $X(t^x_v)$ such that $\mathbf{H}(t^h_u)  \approx \mathbf{H}(t^h_v)$. 

\textit{Stable Unit Treatment Value Assumption:} Denote by $\mathbf{P}$ the set of time-series that are direct predecessors of the effect variable $Y$. Assuming $X \in \mathbf{P}$ (if not, $X$ is independent of $Y$ and therefore there is no causation), the  assumption is violated if $X^{(l)} \in \mathbf{P}$ and $X^{(l)} \not\in \mathbf{H}$, for $ l > 0$. Since units correspond to time-samples, $X^{(l)} \in \mathbf{P}$ implies that the outcome value $Y(t_u^y)$ at time $t_u^y$ depends on the value of the treatment time-series $X$ at time $t_{u-l}^x$. In order to satisfy the  assumption, we modify the $\mathbf{H}$ set as follows:
\begin{equation}
\mathbf{H} := ((X^{(1)}, ..., X^{(L)}) \cap \mathbf{P}) \cup \mathbf{H}, 
\end{equation}
satisfying $Y(t^y_u) \independent X(t^x_v) | \mathbf{H}(t^h_u)$, $\forall u \neq v$.

\textit{i.i.d. Assumption:}  Denote by $Y_1$  the value of the outcome variable for the time-samples that have a positive treatment value and with $Y_0$ for time-samples with zero treatment value. The average causal effect is estimated as $\widehat{E}\{Y_1 - Y_0 | \mathbf{H}\}$. In order to enable statistical inference, the variable $\Delta Y := Y_1 - Y_0 | \mathbf{H}$ needs to be i.i.d.. If $\mathbf{P}$ the set of direct predecessors of $Y$, the outcome value $Y(t_u^y)$ of each time-sample $t_u^y$ will depend on the outcome value $Y(t_{u-l}^y)$ if there is a time-series $Y^{(l)} \in \mathbf{P}$. In case that $Y^{(l)} \notin \mathbf{H}$, the i.i.d. assumption would be violated. In order to satisfy this assumption, we modify the set of time-series $\mathbf{H}$ as follows:

\begin{equation}
\mathbf{H} := ((Y^{(1)}, ..., Y^{(L)}) \cap \mathbf{P}) \cup \mathbf{H}
\end{equation}

Causal inference will be performed by matching on the modified set of time-series $\mathbf{H}$ thus, the variable $\Delta Y := Y_1 - Y_0 | \mathbf{H}$ will be i.i.d.. 

\section{Evaluation}
\label{sec:evaluation}

\subsection{Evaluation with Synthetic Data}
\label{sec:synthetic}
In order to demonstrate the potential of our approach we assess its effectiveness in detecting causal relationships on linear and non-linear synthetic data. We also compare our approach with a multivariate Granger causality model and with an information theoretic approach based on Runge's framework \cite{runge2012escaping} and we demonstrate that our method is more efficient on avoiding false causal conclusions. We denote with $X=\{X(t_u^x): u=1, 2, ... N\}$ and $Y = \{Y(t_u^y): u=1, 2, ... N\}$ the treatment and outcome time-series respectively and with $\mathbf{Z} = \{\mathbf{Z}(t_u^z): u=1, 2, ... N\}$ a set of $M$ confounding variables. We also assume that $t^z_u < t^x_u < t^y_u$, $\forall u$. The relationships among $X$, $Y$ and $\mathbf{Z}$ are described by the following model:

\begin{equation}
X(t_u^x) = h_{xx}(X(t_{u-1}^x)) + f_{xz}(\mathbf{Z}(t_{u}^z)) + \epsilon_x(t_u^x)
\end{equation}
\begin{eqnarray}
Y(t_{u}^y) &=& h_{yy}(Y(t_{u-1}^y)) + f_{yz}(\mathbf{Z}(t_{u}^z)) 
 \nonumber \\
&& + f_{yx}(X(t_u^x))  + \epsilon_y(t_u^y)
\end{eqnarray}
\begin{equation}
Z^i(t_{u}^z) = h_{z^i}(Z^i(t_{u-1}^z)) + \epsilon_{z^i}(t_{u}^z), \forall Z^i \in \mathbf{Z},
\end{equation}
where $\epsilon_x(t_{u}^x)$, $\epsilon_y(t_{u}^y)$ and $\epsilon_{z^i}(t_{u}^z)$ are i.i.d. Gaussian noise variables with zero mean and std. dev. equal to $20 + 2\cdot M$, $10 + 2\cdot M$ and $10$, respectively.

\begin{figure}
  \centering
\includegraphics[width=0.3\textwidth]{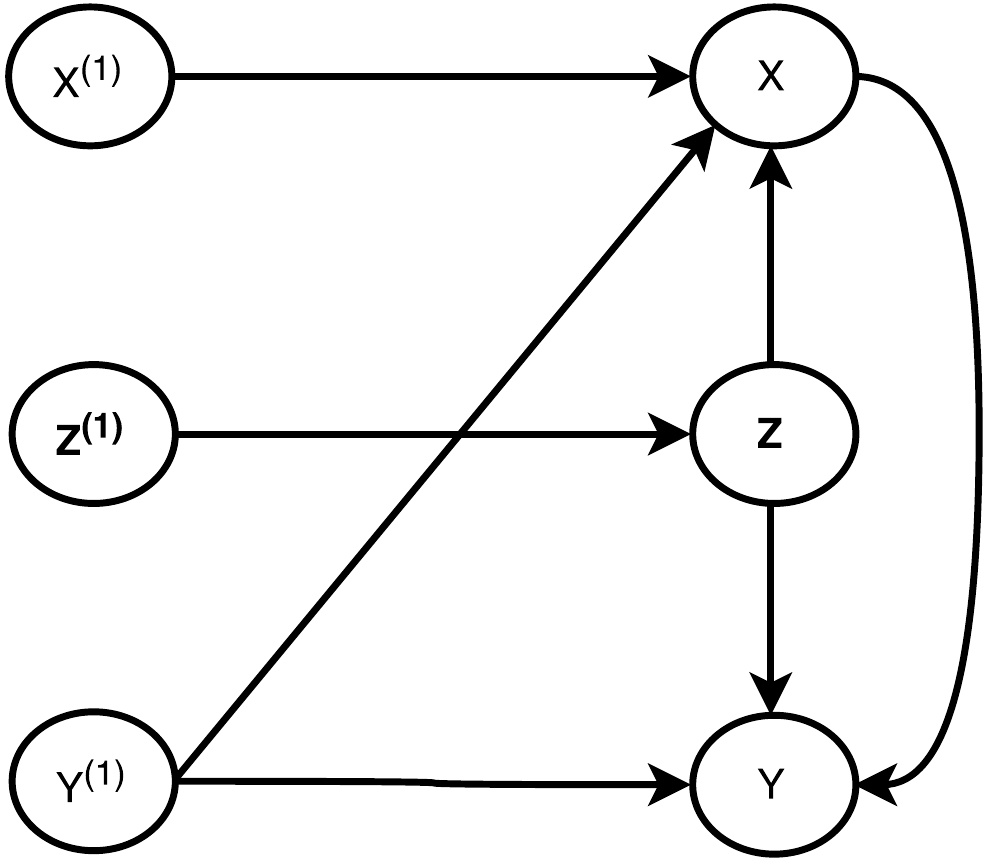}
\caption{Resulting graph after applying Algorithm \ref{Alg:predecessors} on the synthetic data when $M=1$ (i.e. there is only one confounding variable). The graph depicts the direct predecessors of nodes $X$ and $Y$. The set of nodes $\mathbf{H}$ will contain the direct predecessors that nodes $X$ and $Y$ have in common. In the four examined cases $X$ correlates with $Y$, though in Case 2 and Case 4, this is a spurious correlation due to the set of confounding variables $\mathbf{Z}$. There is also a spurious correlation of node $X$ with node $Y^{(1)}$. $X$ and $Y$ are independent to $\mathbf{Z}^{(1)}$ conditional to $\mathbf{Z}$ and $Y$ is independent to $X^{(1)}$ conditional to $X$.}
\label{fig:exampleGraph}
\end{figure}

\begin{figure*}
\centering
       \subfloat[Linear Case.]
                {\includegraphics[width=0.49\textwidth]{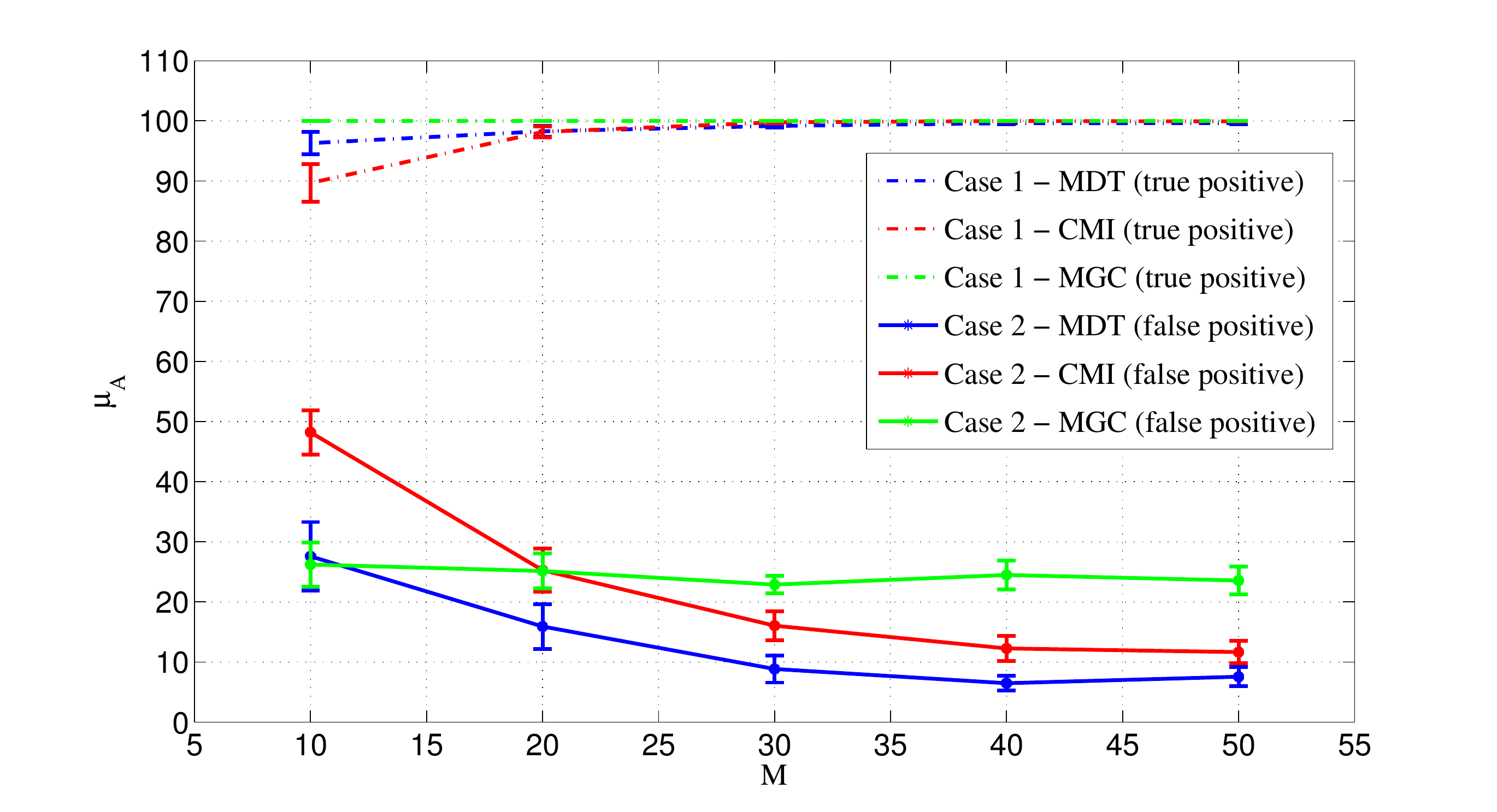}
                \label{fig:linear}}
~
        \subfloat[Non-linear Case.]
                {\includegraphics[width=0.49\textwidth]{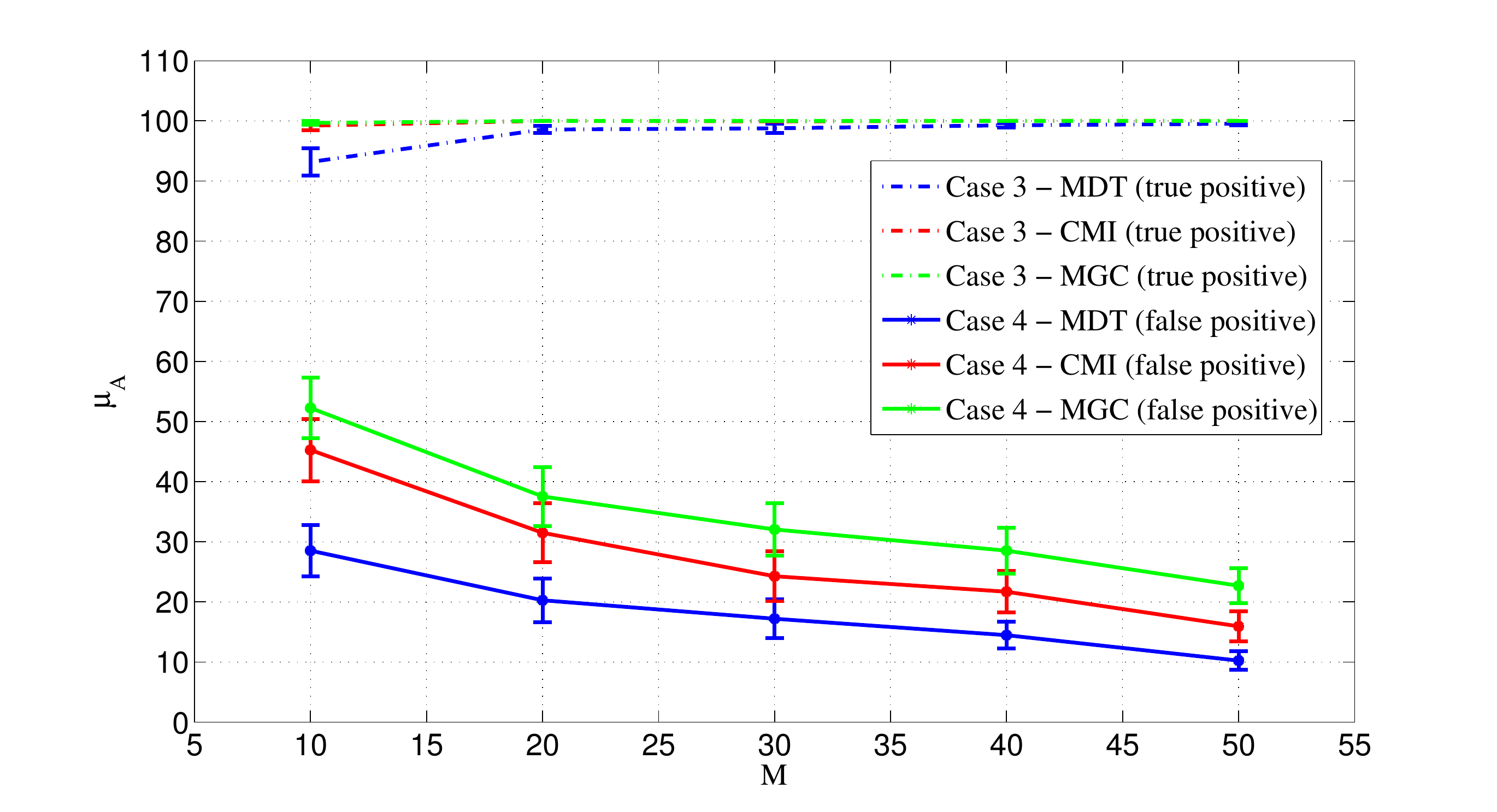}
                \label{fig:noLinear}}

\caption{Comparison of the MDT, CMI and MGC causality detection methods on synthetic data.}
\label{fig:synthetic-res}
\end{figure*}

We consider the following four cases: 

\textbf{Case 1.} The model is linear. Thus, $f_{xz}(\mathbf{Z}(t_{u}^z)) = \sum_i \alpha_{xz,i} \cdot Z^i(t_{u}^z) $, $h_{xx}(X(t_{u-1}^x)) = \alpha_{xx} \cdot X(t_{u-1}^x)$
$h_{yy}(Y(t_{u-1}^y)) = \alpha_{yy} \cdot Y(t_{u-1}^y)$, $f_{yz}(\mathbf{Z}(t_{u}^z)) = \sum_i \alpha_{yz,i} \cdot Z^i(t_{u}^z)$, $f_{yx}(X(t_{u}^x)) = \alpha_{yx} \cdot X(t_{u}^x)$, $h_{z^i}(Z^i(t_{u-1}^z)) = \alpha_{z^i} \cdot Z(t_{u-1}^z)$.

\textbf{Case 2.} We apply the linear model of Case 1, but we set $f_{yx}(X(t_{u}^x)) = 0$. In this case the treatment time-series $X$ does not have any causal impact on the outcome time-series. 

\textbf{Case 3.} The dependences between the confounding variables and the treatment and effect variables are non-linear. In particular,  
\begin{equation*}
f_{xz}(\mathbf{Z}(t_{u}^z)) = \sum_i {\alpha_{xz,i} \cdot (Z^i(t_{u}^z))^2}
\end{equation*}
\begin{equation*}
f_{yz}(\mathbf{Z}(t_{u}^z)) = \sum_i {\alpha_{yz,i} \cdot (Z^i(t_{u}^z))^2} 
\end{equation*}
We use the linear equations of Case 1 for the rest of the functions. 

\textbf{Case 4.} We use the non-linear model of Case 3, but we set $f_{yx}(X(t_{u}^x)) = 0$. In this case, the multivariate linear Granger causality approach may return positive causality result, even though the treatment time-series $X(t)$ does not have any causal impact on the outcome time-series. 

A unit (i.e. time-sample) $u$ of the study is described by the set of time-series values:  $S(t_u):=$ $(X(t_{u}^x)$, $Y(t_{u}^y)$, $\mathbf{Z}(t_{u}^z)$, $X^{(1)}(t_{u}^x)$, $Y^{(1)}(t_{u}^y)$, $\mathbf{Z}^{(1)}(t_{u}^z))$. We apply the following three methodologies on the synthetic data generated using the models above in order to assess the causal impact of variable $X$ on $Y$:

\textbf{Multivariate Granger Causality (MGC).} We apply stepwise regression in order to fit our data to the following model:
\begin{equation}
Y(t_u^y) = a_1 \cdot Y(t_{u-1}^y) + \sum_{l=0}^{(1)} b_l \cdot X(t_{u-l}^x) + \sum_{l=0}^{(1)}  \mathbf{c}_l \cdot \mathbf{Z}(t_{u-l}^z)  + \delta + \epsilon(t_u^y)
\end{equation}
We conclude that $X$ causes $Y$ if $X$ or any lagged version of $X$ is included in the regression model. 

\textbf{Conditional Mutual Information Tests (CMI).} Following Runge's approach \cite{runge2012escaping} a causal graph is created by performing conditional independence tests using conditional mutual information as described at Section \ref{sec:background}.3. 

\textbf{Matching Design for Time-series (MDT).} Following the proposed approach, we apply Algorithm \ref{Alg:predecessors} in order to find the set of variables $\mathbf{H}$ that needs to be controlled in order to achieve conditional ignorability. The resulted graph is depicted at Figure \ref{fig:exampleGraph}. $\mathbf{H}$ includes any $Z^i \in \mathbf{Z}$ that correlates both with $X$ and $Y$.  Moreover, we satisfy the i.i.d assumption by including in $\mathbf{H}$ the time-series $Y^{(1)}$. In order to create groups of treated and untreated units we first transform the time series $X$ into a binary stream $\tilde X$: $\tilde X(t_u^x)=0$, if $X(t_u^x) < \mu_X$;  $\tilde X(t_u^x)=1$, otherwise, where $\mu_X$ is the mean of $X$ (i.e. the $u$-th time-sample corresponds to a treated unit if $X(t_u^x)>\mu_X$). Then, we create pairs of treated and untreated units (i.e. time-samples) by applying Genetic Matching algorithm \cite{diamond2013genetic}. Genetic matching is a multivariate matching method which applies an evolutionary search algorithm in order to find optimal matches which minimize a loss function. We use as a loss function the average standardized mean difference between the treated and control units for all the confounding variables $H^i \in \mathbf{H}$ which is defined as follows:

\begin{equation}
SMD_{H} = \sum_{H^i \in \mathbf{H}} \frac{\sum_{(t_u^h, t_v^h) \in G} |H^i(t_u^h) - H^i(t_v^h)|}{|G| \cdot  \sigma_{H^i}}/|\mathbf{H}|
\label{eq:std-mean}
\end{equation}

Finally, the average treatment is estimated using Equation (\ref{eq:ATE}) and a t-test is used to examine whether the observed $ATE$ is statistically significant from 0.

We generate $100$ samples for each time-series. We vary the number of confounding variables M that are included at set $\mathbf{Z}$ from 10 to 50. In detail, we evaluate the three methodologies for $M = \{10, 20, 30, 40, 50\}$. For each $M$ value, we repeat our study for $30$ randomly selected sets of model coefficients ($\alpha$s). All model coefficients are randomly generated from uniform distribution  on $[-4,4]$ for the linear cases and on $[-1, 1]$ for the non-linear cases. Finally, for each one of the $30$ sets of model coefficients we repeat each study for $100$ different noise realizations. For the $n^{th}$ noise realization, we define:

\begin{equation}
S_n = 
\left\{
  \begin{array}{rcr}
	1 & \mbox{if $X$ was detected as cause of $Y$}\\
    0 & \mbox{otherwise}\\
    
\end{array}
\right.
\end{equation}

For the $k^{th}$ set of model coefficients we also define $A_k = \sum_{n=1}^{100} S_n$. In Case 1 and Case 3, $A_k$ denotes the number of times that a causal relationship from $X$ to $Y$ is successfully inferred (\textit{true positive}) for the $k^{th}$ set of model coefficients and different noise realizations, while in Case 2 and 4 it denotes the number of times that a causal relationship is falsely inferred (\textit{false positive}). In Figure \ref{fig:synthetic-res} we present the mean value of $A_k$, $\mu_A$ along with the standard error of the mean. According to our results, the proposed causal inference technique reduces significantly the number of false positive causality conclusions while it is slightly less successful on detecting real causality for $M=10$. Multivariate Granger causality achieves almost 100\% accuracy on true causality detection both for the linear (Case 1) and non-linear (Case 3) cases. However, it performs poorly in terms of avoiding false positive conclusions. The performance of all the examined methods improves for larger $M$ values (apart from multivariate Granger causality on the linear cases). This is due to the fact that, by adding more variables on the set $\mathbf{Z}$, the dependence of $Y$ and $X$ with each individual $Z^i \in \mathbf{Z}$ is weaker; consequently, although $M$ covariates are used to generate $X$ and $Y$ time-series, for large $M$ values, only a subset of them has significant effect on them. Thus, canceling out the effect of $\mathbf{Z}$ is easier.

\subsection{Causal Effect of Social Media on Stock Markets}
\label{sec:finance}

\subsubsection{Dataset Description}
We apply our method in order to investigate whether information about specific companies and people reactions influence stock market prices. We gather this information by analyzing relevant tweets. Twitter enables us to capture people opinions about the target companies, their optimism/pessimism about stock market movements and their reaction to news such as quarterly results announcements or new product launches. Thus, factors related to the company performance and people trust on the company are reflected on Twitter data. Our study considers the daily closing prices of four big tech companies traded on NASDAQ market: Apple Inc., Microsoft, Amazon.com and Yahoo!. We estimate a daily sentiment index for each of these companies by analyzing the sentiment of related tweets. Tweets are gathered using the the names and the ticker symbols of the examined companies. In detail, we use the hashtags \#apple and \#AAPL for Apple Inc., \#microsoft and \#MSFT for Microsoft, \#amazon and \#AMZN for Amazon.com and \#yahoo and \#YHOO for Yahoo!\footnote{Using the ticker symbol as hashtag for downloading relevant tweets could be problematic for companies with one letter symbol. In such cases, researchers have to identify first the hashtags or names used to identify these companies and search them instead.}. Our study is based on data gathered for four years, from January 2011 to December 2014. We examine whether the sentiment of tweets that are posted before stock market closing time influences the closing prices of the target stocks. In order to eliminate any confounding bias we need to control for factors that may affect both humans sentiment and the target stock prices. Potential influential factor on stocks daily closing prices are their opening prices and their performance during the previous days. Several works have also demonstrated that the performance of other big companies (either local or overseas companies) could influence some stocks (see for example \cite{kenett2010dominating,chi2010network}). Foreign currency exchange rates may also cause money flows to overseas markets and consequently influence stocks prices. Finally, commodities prices could affect the earnings of companies and, therefore, their stocks prices. More specifically, our study involves the following time-series:

 \textbf{The \textit{response} time-series $\mathbf{Y}$.} The difference on the closing prices of the target stocks between two consecutive days. The $u$-th time-sample $t_u^y$ of the time-series $Y$ corresponds to the closing value of the $u$-th day minus the closing value of the previous day. 

 \textbf{The \textit{treatment} time-series $\mathbf{X}$.}  A daily sentiment index that is estimated using tweets related to the target stocks that are posted up to 24 hours before the closing time of the corresponding stock market. In order to assure that the values of the treatment variable are driven by information that has been available before the closing time of the target stocks, we omit from the study tweets posted up to one hour before the closing time. Thus, the sentiment index of day $u$ is estimated using all the tweets posted from 4:00 p.m. (ET) time (i.e. the NASDAQ closing time) of day $u-1$ to 3:00 p.m. (ET) time of day $u$. Consequently, our treatment variable captures the people sentiment and reactions to news realizes at any time during the day, up to one hour before the stock market closing time. Tweets are filtered using the name of the company and the stock symbol as keywords.

 \textbf{The set of time-series $\mathbf{Z}$.} 
 We consider the following time-series which might influence our case study:
\begin{enumerate}
\item \textbf{The difference between the opening and closing prices of two consecutive days.} This time-series is an indicator of the activity of the target stocks at the start of the trading day. 

\item \textbf{The stock market prices of several major companies around the world.} In our study we include all the components of the most important stock market indexes such as NASDAQ-100, Dow-30, Nikkei 225, DAX and FTSE. The study could be influenced only by factors that precede temporally both the treatment and effect variables. Thus, we use the difference between the opening and closing prices of two consecutive days for stocks that are traded in the USA exchange markets. The closing time of companies traded at the overseas markets precedes the closing time of the USA stock exchange market, thus the time-series for all the overseas companies stocks correspond to the difference on the closing prices between two consecutive days. Although the values of the treatment variable are driven by tweets that are posted both before and after the corresponding values of the time-series that we use to describe the performance of big companies, for convenience, we consider that the $u$-th time-sample $t_u^x$ of the treatment time-series occurs one hour before the USA stock exchange market closing time at day $u$.  Thus, the time-sample $t_u^z$ of any of the time-series that are used to describe the performance of either a USA-based company or an overseas company temporally precedes the $u$-th sample of the treatment time-series.

\item \textbf{The daily opening values of foreign currency exchange rates minus the previous day opening values.} We include the exchange rates between dollar and British pound, Euro, Australian dollar, Japanese Yen, Swiss Franc and Chinese Yen. 

\item \textbf{The difference between the opening values of commodities for consecutive days.} We include the following commodities: gold, silver, copper, gas and oil. 
\end{enumerate}

\subsubsection{Daily Sentiment Index Estimation}

We classify each tweet as negative, neutral or positive using the SentiStrength classifier~\cite{thelwall2013heart}. SentiStrength estimates the sentiment of a sentence using a list of terms where each term is assigned a weight indicating its positivity or negativity. We updated the list of terms in order to include terms that are commonly used in finance \footnote{The words list along with the updated words weights can be found here: https://www.cs.bham.ac.uk/~tkt357/Publications.html}. 

Sentiment extraction from text may be inaccurate. Although this issue has been disregarded in previous works \cite{bollen2011twitter, zhang2012predicting, zhang2011predicting}, here, in order to account for such inaccuracies on sentiment classification, we estimate a probability distribution function of the daily sentiment instead of a single metric. Let us define a set of three objects $S = \{positive, neutral, negative\}$. Each object $i \in S$ denotes a classification category. Let us also define a random variable $V_i$ as follows:

\begin{equation}
V_i = 
\left\{
  \begin{array}{rcr}
    0 & \mbox{if a negative tweet is classified in class i}\\
    1 & \mbox{if a neutral tweet is classified in class i}\\
    2 & \mbox{if a positive tweet is classified in class i}\\
\end{array}
\right.
\end{equation}

We derive the probability distribution functions of each random variable $V_i$, with $i \in S$, based on the classification performance results. We evaluate the performance of the classifier by manually classifying 1200 randomly selected tweets (200 tweets for each one of the four examined companies). The probability distribution functions are presented in Table \ref{tab:accuracy}.

\begin{table}[t]
\centering
\caption{Accuracy of the text classification for each classification category.}
\begin{tabular}{ |c|c|c|c| }
  \hline 
  {}&{$P(V_i = 0)$}&{$P(V_i = 1)$}&{$P(V_i = 2)$}\\
  \hline 
  {i = positive}&{0.05}&{0.27}&{0.68}\\
  \hline 
  {i = neutral}&{0.03}&{0.91}&{0.06}\\
  \hline 
  {i = negative}&{0.65}&{0.29}&{0.06}\\
  \hline 
\end{tabular}
\label{tab:accuracy}
\end{table}

Let us define with $N_i$ the number of tweets posted within a day that are classified in category $i$. We define a random variable $\mathcal{V}_{t_u}$ which corresponds to the sentiment of the $u$-th day as follows:

\begin{equation}
\mathcal{V}_{t_u} = \sum_{i\in S} N_i \cdot V_i 
\end{equation}

Since $2$ is the maximum value of $V_i$, $\mathcal{V}_{t_u} \in \{0, 1, ..., 2 \cdot \sum_i N_i \}$. We estimate the probability distribution of $\mathcal{V}_{t_u}$ by deriving the probability-generating function under the assumption that the real sentiment of a tweet is independent to the sentiment of any other tweet conditional to the observed classification of the tweet sentiment (i.e. the inferred sentiment by SentiStrength). Although the sentiment of a tweet may depend on previously posted tweets, given that the probability of correctly inferring the sentiment of a tweet is independent to the sentiment inference of any other tweet, our assumption is realistic. The probability-generating function of $\mathcal{V}_{t_u}$ is expressed as follows:

\begin{equation}
G_{\mathcal{V}_{t_u}} = \prod_{i\in S} (G_{V_i}(z))^{N_i} = \prod_{i\in S} (\sum_{x=0}^{2} p(V_i = x)\cdot z^x)^{N_i}
\end{equation}

The probability distribution function of $\mathcal{V}_{t_u}$ is estimated by taking the derivatives of $G_{\mathcal{V}_{t_u}}$. If $\mathcal{N}_{t_u}$ the number of tweets posted the  $u$-th day, then, $\mathcal{V}_{t_u} \in \{0, 1, ..., \mathcal{N}_{t_u} \cdot M\}$ and the probability that the general sentiment of the $u$-th day is positive is given by the probability $P_{pos}(t_u) = P(\mathcal{V}_{t_u}>\frac{\mathcal{N}_{t_u} \cdot M}{2})$. 

\subsubsection{Results}
\label{sec:results}
\begin{figure}
  \centering
\includegraphics[width=0.47\textwidth]{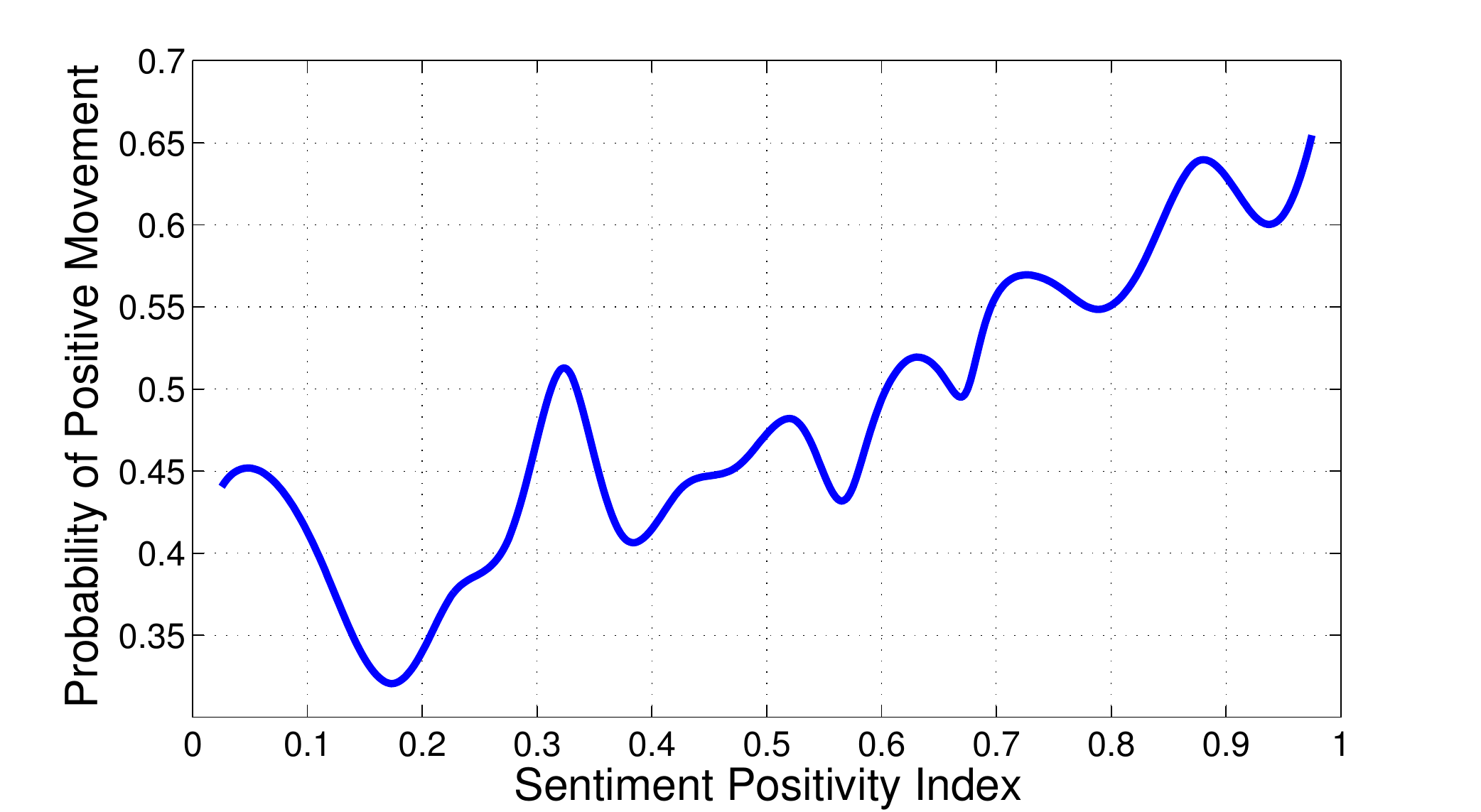}

\caption{Probability distribution function of having a positive movement on the traded assets prices conditional to the sentiment of the tweets.}
\label{fig:probPosMovement}
\end{figure}

\begin{figure*}
  \centering
                \includegraphics[width=0.75\textwidth]{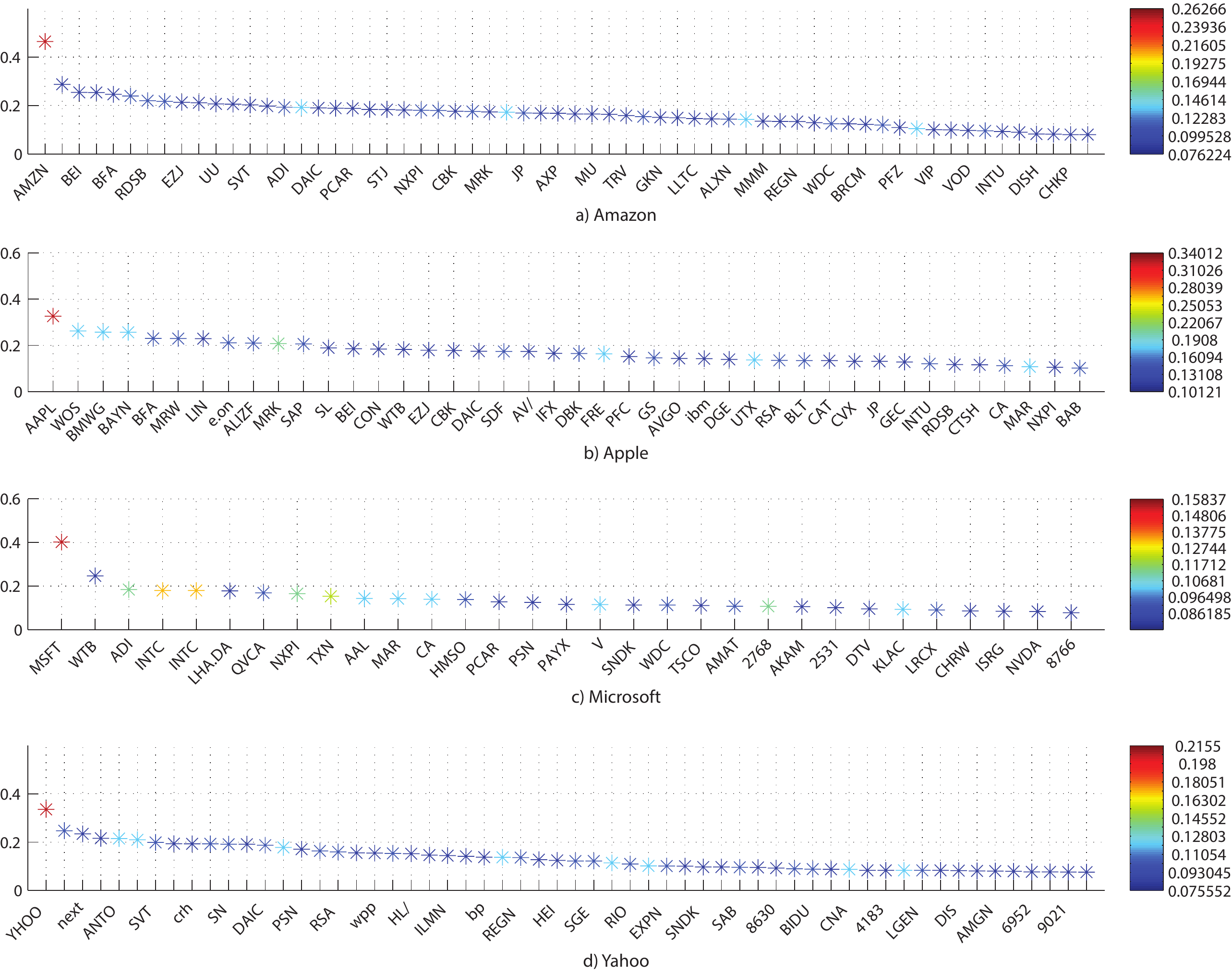}
\captionsetup{justification=centering}
\caption{Correlation between the confounding variables and the treatment and effect time-series.}
\label{fig:correlations}
\end{figure*}

We create a binary treatment variable $X$ by applying thresholds on $P_{pos}(t_u)$. More specifically, a unit that describes the $u$-th day of the study is considered to be treated (i.e. $X(t_u^x)=1$) if $P_{pos}(t_u^x) \geq P_{thresh}^1$ and untreated (i.e. $X(t_u^x)=0$) if $P_{pos}(t_u) < P_{thresh}^0$. We conduct our study for three different pairs of thresholds. In detail, we consider a pair of thresholds $T1$, where thresholds $P_{thresh}^1$ and $P_{thresh}^0$ are set to the $50^{th}$ percentile of $X$, a pair of thresholds $T2$ where $P_{thresh}^1$ is set to the $60^{th}$ percentile of $X$ and $P_{thresh}^0$ to the $40^{th}$ percentile of $X$ and finally a pair $T3$ where $P_{thresh}^1$ and $P_{thresh}^0$ are set to the $70^{th}$ and $30^{th}$ percentiles respectively. By increasing the value of $P_{thresh}^1$ and decreasing the value of $P_{thresh}^0$ we eliminate from our study days in which the estimated tweets polarity is uncertain either due to measurement error or because the overall sentiment that is expressed during these days is considered to be neutral. Although discretization of a continuous variable results in information loss which may jeopardize, in some cases, the reliability of the causal inference, we enhance the validity of our conclusions by considering different threshold values. 

We include in our study all the previously mentioned variables. We found that there is no autocorrelation in our time-series, thus, since there is no dependence of our time-series on their past values, we set the maximum lag $L$ equal to 1. For each of the four target stocks, we applied Algorithm \ref{Alg:predecessors} in order to find the set of time-series $\mathbf{H}$ that needs to be controlled. We consider a correlation to be statistically significant if the corresponding p-value is smaller than 0.05. We used Spearman's rank correlation in order to capture potentially non-linear relationships among the examined variables. We found that stock movements are significantly correlated with the sentiment of tweets posted within the same day. Our findings are in agreement with results of other studies \cite{bollen2011twitter, preis2010complex, zhang2011predicting}. We also found that stock prices are independent to past tweets sentiment conditional to more recent tweets. This indicates that any effect of tweets on stock prices is instant rather than long-term. Finally, according to our results, the daily movement of the traded assets for the target companies does not correlate with past days movements. This finding is consistent with the weak-form efficient market hypothesis according to which, it is not feasible to predict stock market movements by applying technical analysis. In Table~\ref{tab:correlation} we present the correlation coefficient of the effect variable $Y$ with the treatment variable $X$ and the 1-lagged variables $X^{(1)}$ and $Y^{(1)}$ for each one of the four examined companies. In Figure \ref{fig:probPosMovement} we present the empirical probability distribution function of having a positive movement on the traded assets prices conditional to the sentiment of the tweets $P(Y(t_u^y) >0 | X(t_u^x))$. The probability distribution function is estimated using data collectively for the four examined companies. Our results indicate that the probability of having a positive movement on the stock market does not increase linearly with the daily tweets positivity index. Stock market movement is quite uncertain when the positivity index of the tweets ranges between 0.35 and 0.65, while the probability of having a positive movement is increasing for positivity index larger than 0.65. Moreover, we notice a relatively high probability of having a positive movement in days with sentiment positivity index lower than 0.1. Considering that daily tweets sentiment capture the current and past stock market trends, this could be attributed to the fact  that investors may consider that it is a good time to invest money when assets prices are low; consequently, this could give lead to an increase of stock market prices.

\begin{table}[th]
\centering
\caption{Correlation of $Y$ with $X$, $X^{(1)}$ and $Y^{(1)}$.}
\begin{tabular}{ |c|c|c|c|c|}
  \hline 
  {}&{AAPL}&{MSFT}&{AMZN}&{YHOO}\\
  \hline 
  {$X$}&{0.393}&{0.155}&{0.237}&{0.273}\\
  \hline
 {$X^{(1)}$}&{ 0.032}&{0.036}&{0.012}&{0.046}\\
\hline
 {$Y^{(1)}$}&{0.009}&{ -0.003}&{-0.037}&{0.031}\\
  \hline  
\end{tabular}
\label{tab:correlation}
\end{table}

Moreover, we find that both the effect and the treatment variables correlate with the most recent stock prices of several local and overseas companies. The daily movements of the target stocks correlate with US dollar exchange rates; however, currency exchange rates do not have any impact on the treatment variable. In Table \ref{tab:number-of-confounders} we present the number of variables from each category that will be included in the set $\mathbf{H}$ for the four target companies and in Figure \ref{fig:correlations}, we present the correlation coefficients of the treatment and effect time-series with all variables in set $\mathbf{H}$. For all the examined stocks, the strongest confounder is their opening prices.

\begin{table}[th]
\centering
\caption{Number of variables that are included in the set $\mathbf{H}$ for each of the four examined companies.}
\begin{tabular}{ |c|c|c|c|c|}
  \hline 
  {}&{AAPL}&{MSFT}&{AMZN}&{YHOO}\\
  \hline 
  {Nasdaq-100 Comp.}&{6}&{21}&{33}&{7}\\
  \hline
 {Nikkei Comp.}&{1}&{3}&{1}&{13}\\
  \hline 
{DAX Comp.}&{18}&{2}&{7}&{10}\\
  \hline 
{FTSE Comp.}&{10}&{3}&{12}&{26}\\
  \hline 
{Dow-30 Comp.}&{7}&{3}&{9}&{2}\\
  \hline 
{FOREX}&{0}&{0}&{0}&{0}\\
  \hline  
{Commodities}&{0}&{0}&{0}&{0}\\
  \hline  

\end{tabular}
\label{tab:number-of-confounders}
\end{table}

\begin{figure}
  \centering
\includegraphics[width=0.47\textwidth]{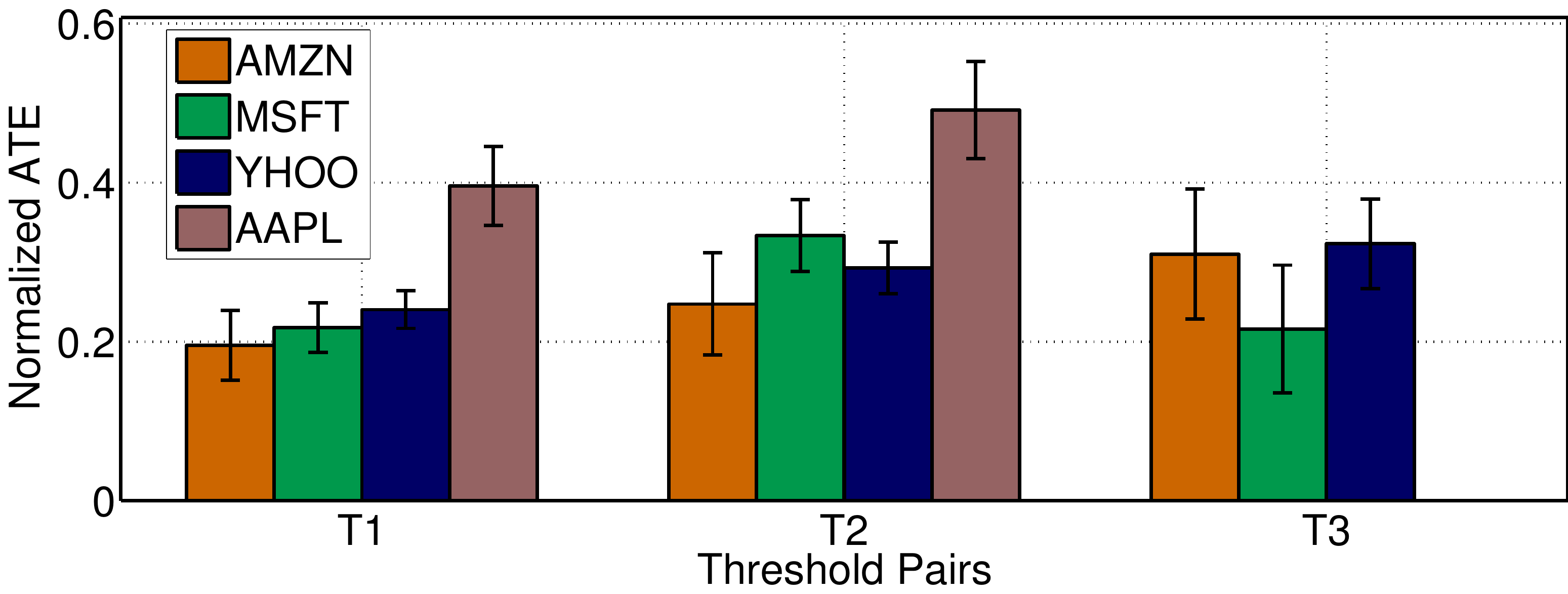}
\caption{Normalized ATE for the three threshold pairs.}
\label{fig:ATE}
\end{figure}

\begin{figure}
  \centering
\includegraphics[width=0.47\textwidth]{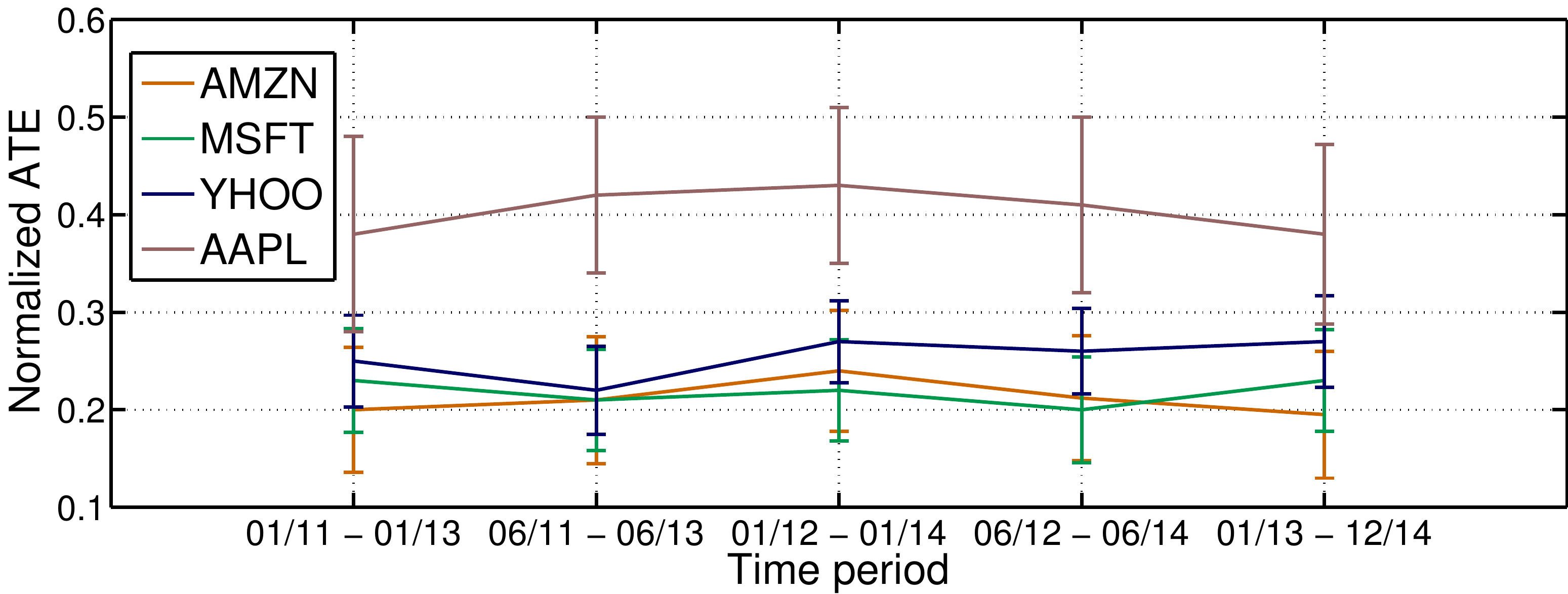}
\caption{Normalized ATE for the threshold pair T1. Analysis is conducted in two-year sub-periods using sliding windows with 6 months step.}
\label{fig:subperiods}
\end{figure}

In order to eliminate the effect of the confounding variables we need to match treated and control units with similar values on their set of confounding variables. We create optimal pairs of treated and untreated units by applying Genetic Matching algorithm \cite{diamond2013genetic}, using as loss function the average standardized mean difference between the treated and control units, as described in Equation \eqref{eq:std-mean}, for all the confounding variables. We check if sufficient balance between treated and untreated subjects has been achieved by checking the standardized mean difference for each confounding variable. The remaining bias from a confounding variable is considered to be insignificant if the standardized mean difference is smaller than $0.1$ \cite{austin2009relative, austin2008goodness}. 

\begin{figure*}[t!]
\centering
       \subfloat[Amazon.]
                {\includegraphics[width=0.4\textwidth]{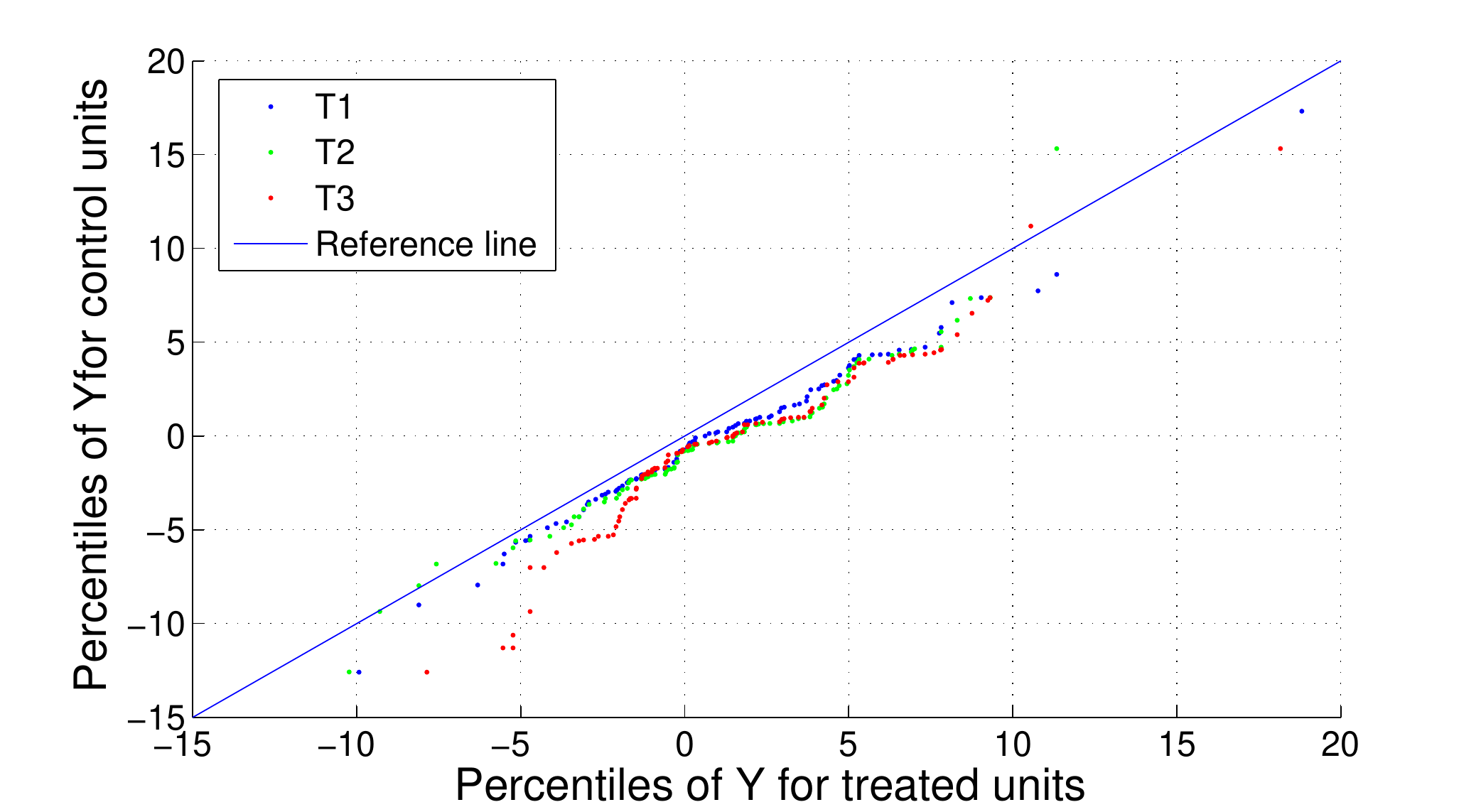}
                \label{fig:amazon}}
~
        \subfloat[Apple.]
                {\includegraphics[width=0.4\textwidth]{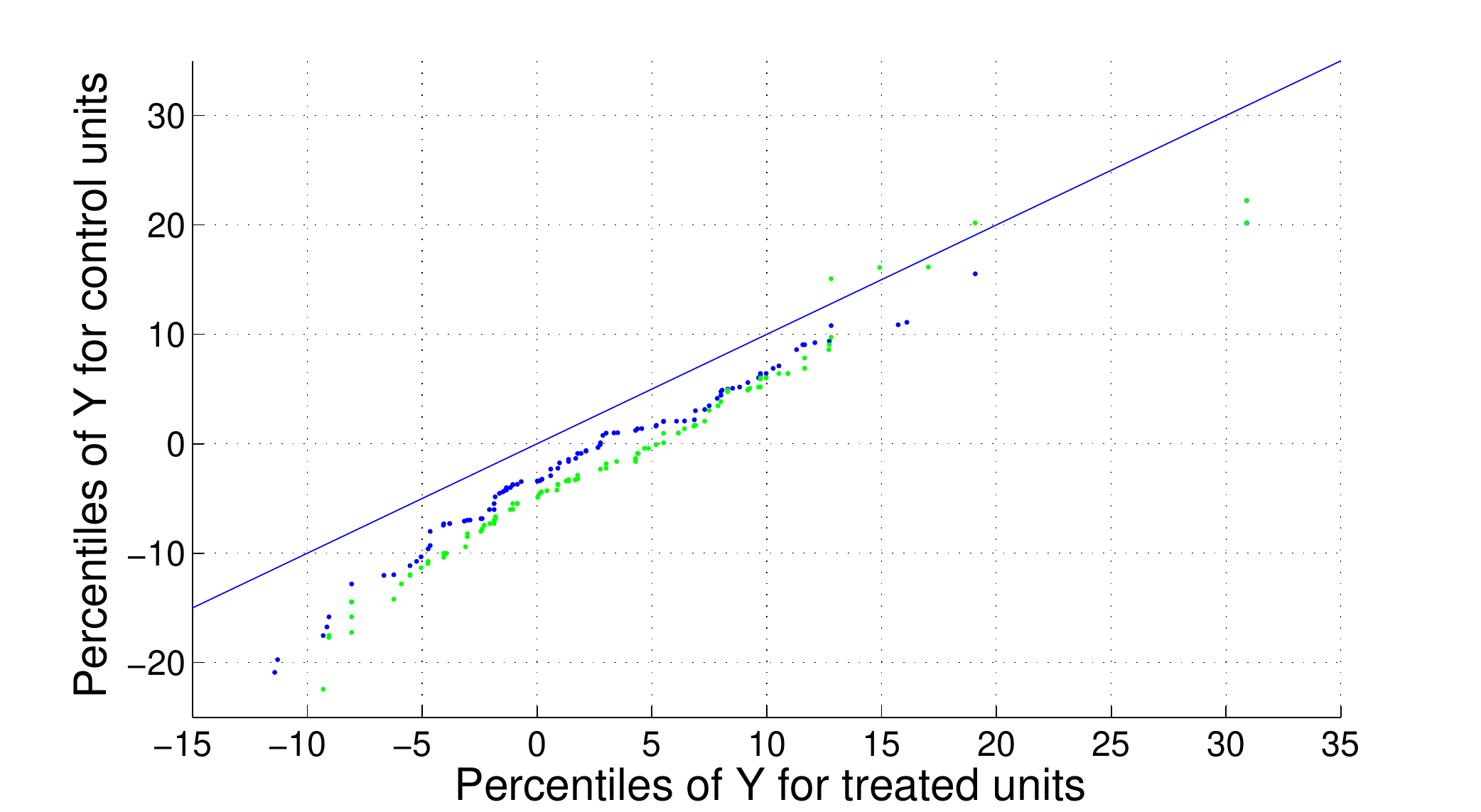}
                \label{fig:apple}}
\\
        \subfloat[Microsoft.]
                {\includegraphics[width=0.4\textwidth]{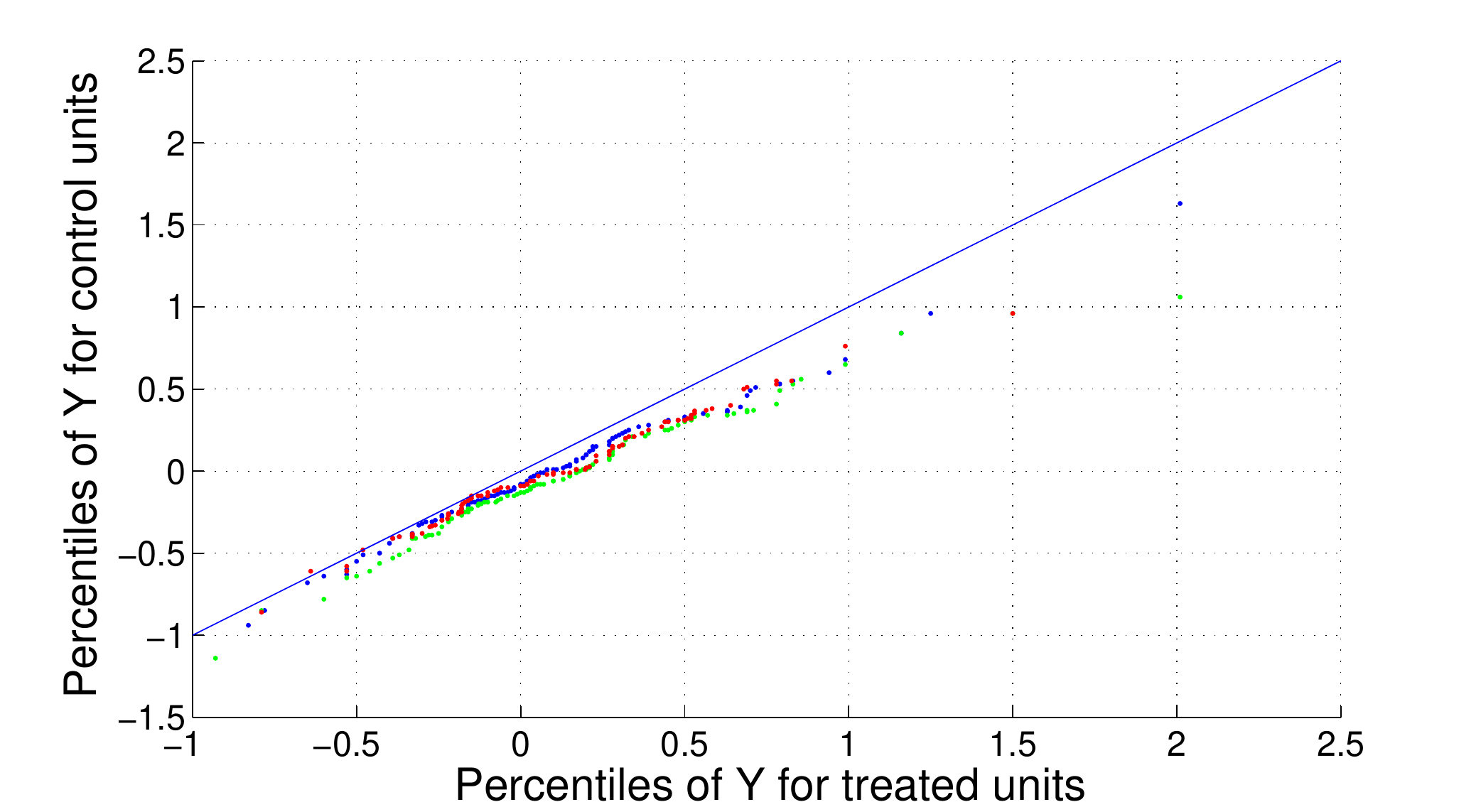}
                \label{fig:msft}}
~
        \subfloat[Yahoo!.]
                {\includegraphics[width=0.4\textwidth]{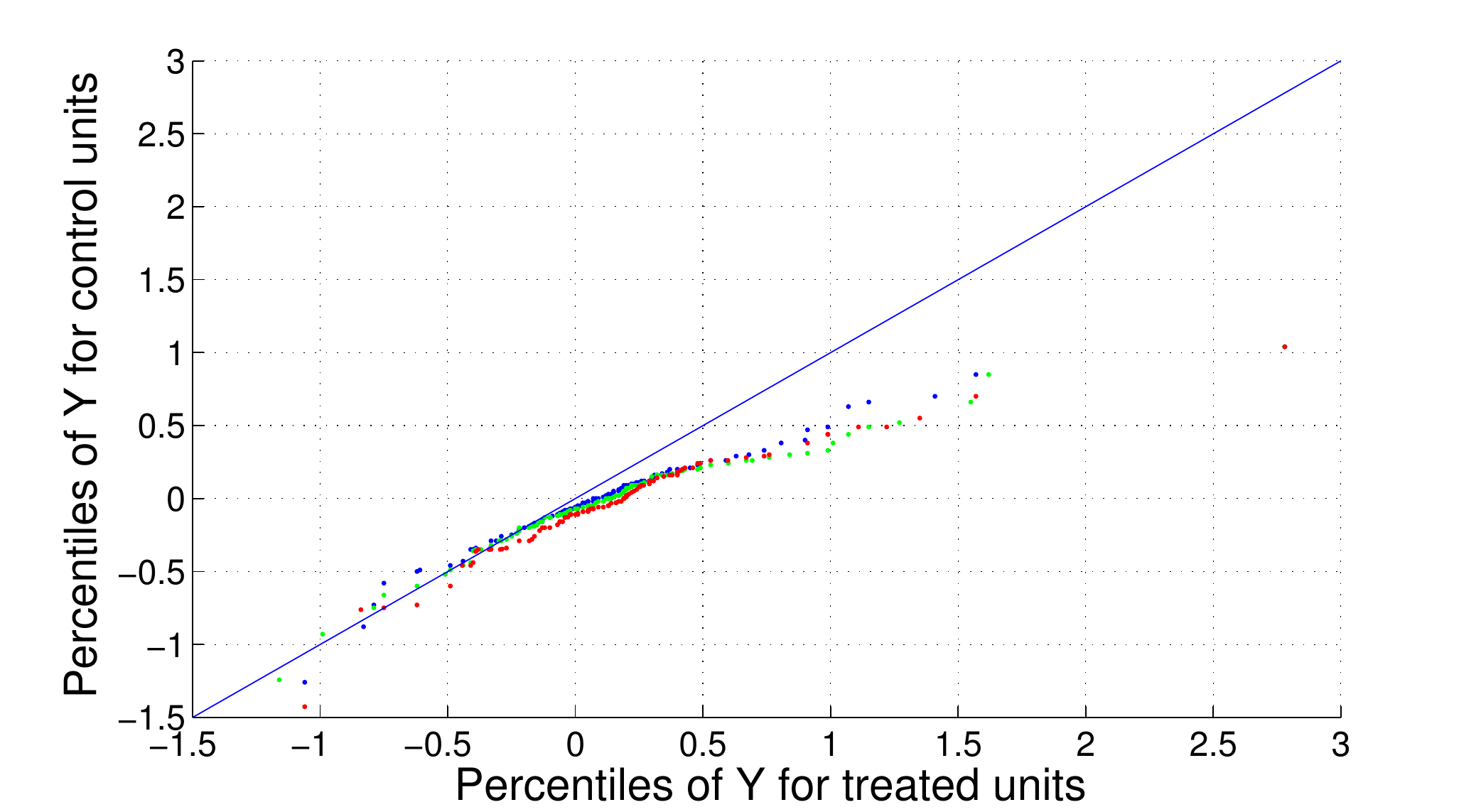}
                \label{fig:yahoo}}
\caption{Percentiles of treated units versus percentiles of matched control units.}
\label{fig:qqplot}
\end{figure*}

We examine the causal effect of the sentiment of tweets on the target stocks for the three pairs of thresholds. We apply Equation \ref{eq:ATE} in order to estimate the average treatment effect (ATE). Under the assumption that the examined treatment has no impact on the effect variable, the ATE would be equal to 0. We use a t-test to assess how significant is the difference of the observed ATE value from 0. In Figure \ref{fig:ATE}, we present the average treatment effect normalized by the variance of the effect variable $Y$ along with the 95\% confidence interval values. According to our results, the effect of the tweets sentiment on the stocks prices of all the examined stocks is statistically significant. We also observe that the causal impact is stronger for larger values of the $P_{thresh}^1$ and smaller $P_{thresh}^0$ threshold values, i.e. the observed difference on the effect variable between the treatment and control groups is larger when we consider only days for which there is less uncertainty on the estimated tweets polarity. For Apple, it was not possible to create balanced treated and control groups for the thresholds pair $T3$. This is due to the fact that the opening prices of the AAPL stocks are very strongly correlated with both the effect and treatment variables and, therefore, there were not enough treatment and control units with similar values on their confounding variables. Since any causal conclusions are not reliable when the treated and control groups are not balanced, we do not present results for Apple for this pair of threshold values. In addition, we repeat our study for different time-periods using a two-year sliding window with six-month step. In Figure \ref{fig:subperiods} we present our findings for the four examined companies and the first pair of thresholds. According to our results, the difference on the estimated ATE is insignificant for the examined sub-periods.  
Finally, in Figure~\ref{fig:qqplot} we compare the distributions of the effect variable $Y$ for the treated and control units by plotting their percentiles against each other. Under the hypothesis that the treatment variable has no effect on variable $Y$, the plot should follow approximately the line $y=x$. However, most of the points of the plot lie below the reference line $y=x$, indicating that the majority of the percentiles of variable $Y$ for the treated units are larger than the corresponding percentiles for the control units. 

Finally, we examine the impact of time-series $P_{pos}$ on $Y$ by applying multivariate Granger causality (MGC) and conditional mutual information tests (CMI) as described at Section \ref{sec:synthetic}. In Table \ref{tab:otherMethodsTwitter} we present the p-values for the two methods under the null hypothesis that the effect of $P_{pos}$ on $Y$ is zero. According to our results, both methods reject the null hypothesis with p-value smaller than $0.05$ and confirm our findings that there is a causal link between social media and traded assets prices for the four examined companies. Thus, in this case study, all the examined causal inference methods result in the same conclusion.  

\begin{table}[th]
\centering
\caption{Resulted p-values for the methods MGC and CMI under the null hypothesis that the influence of $P_{pos}$ on $Y$ is zero.}
\begin{tabular}{ |c|c|c|c|c|}
  \hline 
  {}&{AAPL}&{MSFT}&{AMZN}&{YHOO}\\
  \hline 
  {MGC}&{0.0000}&{0.0008}&{0.0000}&{0.0011}\\
  \hline
 {CMI}&{ 0.0002}&{0.0021}&{0.0001}&{0.0004}\\
\hline
\end{tabular}
\label{tab:otherMethodsTwitter}
\end{table}

\subsubsection{Sensitivity Analysis}

The main limitation of all non-experimental causality studies is that they are based on the assumption that all confounding variables are known. However, in real scenarios there may be unmeasured factors which influence the assignment of units to treatments. In such cases, the conditional ignorability assumption is violated and consequently, any causal inference result may be biased. In our study, we include a large number of potentially influential factors such as the performance of other companies traded assets, commodities prices and currency exchange rates. However, there are other factors, such as inflation rates, political changes or economic policy changes which could influence both people sentiment, captured through Twitter, and traded assets prices. Although such factors may be reflected on the observed confounding variables (e.g. macroeconomic factors such as inflation rates would also affect the prices of other traded assets and consequently, the observed Twitter sentiment may be independent to inflation rates conditional to the performance of other assets included in the study), there may still be some bias due to unobserved factors.  

A sensitivity analysis can be conducted in order to assess how the results of the study would be influenced in the presence of unmeasured confounding variables \cite{rosenbaum2002observational, rosenbaum1983assessing}. In detail, let us denote with $\pi_{t_u}$ the probability that a unit $t_u$ corresponding to the $u$-th day is assigned to a treatment (i.e. $X(t_u^x) = 1$) and $O_{t_u} = \pi_{t_u} / (1 - \pi_{t_u})$ the odds of the unit to receive a treatment. Then, we denote with $\Gamma = O_{t_u} / O_{t_v}$ the ratio of the odds of two units $t_u$, $t_v$. If $\Gamma = n$, the unit $t_u$ is $n$ times more likely to receive a treatment than unit $t_v$ due to unobserved factors. Under the conditional ignorability assumption (i.e. units are equally likely to receive a treatment conditional to their observed characteristics), the ratio $\Gamma$ should be equal to $1$ for two matched time-samples $t_u$ and $t_v$. 

In \cite{rosenbaum2002observational}, Rosenbaum applies the Wilcoxon's signed rank test \cite{lehmann1975nonparametrics} for the resulted matched treated and control pairs of a causality study under the null hypothesis that the treatment has no effect on the observed outcome variable. According to this method, for each matched pair $(t_u, t_v)$ a rank is assigned to the outcome difference $Y(t_u^y) - Y(t_v^y)$. The Wilcoxon's signed rank statistic $W$ is estimated as the sum of the ranks of the positive differences (the interested reader can find a detail description of the method in \cite{lehmann1975nonparametrics}). Under the null hypothesis, the mean value of $W$ is $S \cdot (S + 1) /4$, where $S$ the number of matched samples. When $S$ is sufficiently large, the upper bound of the distribution of $W$ can be approximated by a normal distribution with mean $􏰀􏰀\Gamma/ (1 + 􏰀\Gamma) \cdot S \cdot (S + 1) /2$. Thus, the sensitivity on unobserved confounding variables can be assessed by computing the upper bounds on the p-values of the Wilcoxon's signed rank test for increasing $\Gamma$ values.

We apply this method in order to evaluate the sensitivity of our results on unobserved confounding variables. At Table \ref{tab:sensitivityAnalysis}, we present the results of our sensitivity analysis for $\Gamma \leq 2.0$ and for the $T2$ pair of thresholds. According to our results, the causal influence of Twitter on Apple stock prices would be considered statistically significant (with p-value 0.014) even if some days were twice more likely (i.e. $\Gamma = 2$) to have positive sentiment conditional to the observed confounding variables due to unmeasured factors. Similarly, for Amazon our causal inference results are statistically significant (i.e. p-value $< 0.05$) for $\Gamma \leq 1.9$, for Yahoo! for $\Gamma \leq 1.7$ and finally for Microsoft our conclusion would be invalid for $\Gamma \geq 1.6$. 

\begin{table}[th]
\centering
\caption{Sensitivity Analysis.}
\renewcommand{\arraystretch}{1.25}
\begin{tabular}{|c|c|c|c|c|}
	\hline
  \multirow{2}{*}{$\mathbf{\Gamma}$} &\multicolumn{4}{c|}{\textbf{Upper bound on p-value}}
  \\\cline{2-5}
  {}&{AAPL}&{AMZN}&{YHOO}&{MSFT}\\
  \hline 
  {1.0}&{0.0000}&{0.0000}&{0.0000}&{0.0000}\\
  \hline 
  {1.1}&{0.0000}&{0.0000}&{0.0000}&{0.0000}\\
  \hline 
  {1.2}&{0.0000}&{0.0000}&{0.0000}&{0.0005}\\
  \hline 
  {1.3}&{0.0000}&{0.0002}&{0.0000}&{0.0027}\\
  \hline 
  {1.4}&{0.0001}&{0.0011}&{0.0003}&{0.0109}\\
  \hline 
  {1.5}&{0.0003}&{0.0042}&{0.0010}&{0.0327}\\
  \hline 
  {1.6}&{0.0009}&{0.0131}&{0.0034}&{0.0775}\\
  \hline 
  {1.7}&{0.0021}&{0.0328}&{0.0091}&{0.1519}\\
  \hline 
  {1.8}&{0.0043}&{0.0692}&{0.0209}&{0.2552}\\
  \hline 
  {1.9}&{0.0082}&{0.1263}&{0.0419}&{0.3789}\\
  \hline 
  {2.0}&{0.0142}&{0.2050}&{0.0749}&{0.5093}\\
  \hline 
\end{tabular}
\label{tab:sensitivityAnalysis}
\end{table}

\section{Related Work}
\label{sec:related-work}

Several works have previously examined the potential of information extracted from social media, search engine query data or other web-related information to predict stock market returns. For example, in \cite{zhang2011predicting} the authors demonstrate that the level of optimism/pessimism, which is estimated using Twitter data, correlates with stock market movement. 
Other projects have been focussed on the possible use of sentiment analysis based on Twitter data for the prediction of traded assets prices by applying a bivariate Granger causality analysis~\cite{bollen2011twitter, zhang2012predicting} or regression models \cite{si2013exploiting}.
Similarly, in \cite{zheludev2014can} information theoretic methods are used to investigate whether sentiment analysis of social media can provide statistically significant information for the prediction of stock markets and in \cite{porshnev2013machine} authors propose a prediction method based on machine learning. 
In \cite{preis2010complex, preis2013quantifying} the authors have also demonstrated that search engine query data correlate with stock market movements. In \cite{moat2013quantifying} the authors propose a trading strategy that utilizes information about Wikipedia views. They demonstrate that their trading strategy outperforms random strategy. However, all the above mentioned studies are based on bivariate models. Although their results indicate that social media and other web sources may carry useful information for stock market prediction, by using these techniques it is not possible to figure out if other factors are influencing the observed trends. 

Trading strategies that utilize both technical analysis and sentiment analysis are discussed in \cite{schumaker2009textual, deng2011combining}. However, these works are based on regression analysis, thus they suffer from the limitations that have been previously discussed. Moreover, all the studies so far focus mainly on prediction of stock market movement. Although they provide insights about the influence that emotional and social factors may have on stock market, they do not investigate the presence of causality. To the best of our knowledge, this is the first work that attempts to measure the causal effect of such factors on stock markets.

\section{Discussion}
\label{sec:discussion}

Our results on the simulated experiments indicate that our method is more effective on avoiding false positive causality conclusions. We have examined the performance of the proposed method in datasets with up to 50 dimensions and 100 time-samples. Extreme high-dimensional cases with $p > n$ are not considered in this study. In such cases, balancing treatment and control groups for each confounding variable would require large number of samples. \emph{Propensity score matching} \cite{harder2010propensity} represents an alternative matching method that can effectively handle large number of confounding variables by performing matching on a single balancing score, i.e. the \emph{propensity score}. The propensity score corresponds to the probability of a unit to be assigned to a treatment and it is usually approximated by applying a logistic regression model of the treatment against the set of confounding variables. \emph{High-dimensional propensity score matching} \cite{schneeweiss2009high} has also been proposed in order to handle extreme high-dimensional cases with $p >> n$.

One of the main advantages of the proposed MDT method over multivariate Granger causality and CMI is that the design of the study is separated from the analysis. The values of the response time-series $Y$ are not used during the matching process. The causal impact of a time-series $X$ on $Y$ is evaluated only after sufficient balance between the treated and untreated samples has been achieved. In contrast, in a regression-based analysis the response time-series $Y$ is used in order to learn the coefficients of the predictor variables of the study. Many studies suggest that regression-based methods for causal inference are less reliable \cite{austin2011introduction}. Moreover, the proposed method is non-parametric, while Granger causality is based on assumptions about the model class (i.e. linear/non-linear relationships). According to our results, linear Granger causality performs poorly when there are non-linear relationships among the examined time-series. 

In addition, as it was previously discussed, MDT requires significantly fewer conditional independence tests and smaller conditioning sets. In detail, the maximum conditioning set of the proposed method is equal to the maximum lag $L$, while the maximum conditioning set of CMI is $M \cdot L$ (with M the number of confounding variables). Thus, MDT can handle more effectively datasets which include large number of confounding variables. 

Moreover, the computational cost of creating the graph is significantly lower for MDT compared to CMI. Assuming discrete time-series with values in a set $V$, the computational complexity of creating a graph by applying the proposed method is $O(|V|^L \cdot M \cdot N)$, while the computational cost of CMI is $O(|V|^{M \cdot L} \cdot M \cdot N)$, with $|V|$ the size of set $V$. However, causal inference with MDT requires an additional matching step, and consequently, its computational complexity largely depends on the matching method that is applied. If a simple nearest neighbor matching method is applied \cite{MatchingSurvey}, the cost of finding the best match of a single unit is $O(|\mathbf{H}|\cdot N)$, with $|\mathbf{H}|$ denoting the size of set $\mathbf{H}$, and the cost of matching all the units is $O(|\mathbf{H}|\cdot N^2)$. Thus, the overall computational cost of MDT is $O(|V|^L \cdot N + |\mathbf{H}| \cdot N^2)$. However, when more complex matching methods, such as Genetic matching \cite{diamond2013genetic}, are applied, the computational cost of MDT can be significantly larger. Genetic matching algorithm applies an evolutionary search method in order to find optimal weights for each covariate. In each algorithm iteration, a set of $P$ weights for each one of the variables in set $\mathbf{H}$ is generated. $P$ corresponds to the \emph{population size} of the genetic algorithm. Then, nearest neighbor matching is applied on the weighted variables of $\mathbf{H}$ for each one of the $P$ weights. A loss function is used to estimate the loss for each of the $P$ resulted sets of matched pairs. If the loss is sufficiently small for any of the $P$ weights, the method terminates; otherwise this process is repeated. A maximum number of iterations $I$ can be used in order to set an upper bound on the computational time of the method. In our experiments, we used as loss function the average standardized mean difference between the treated and control units. The cost of loss estimation for each weights set is $O(|\mathbf{H}|\cdot N)$. Thus, the total computational cost of the matching process is $O(I\cdot P\cdot |\mathbf{H}| \cdot N^2)$. Although the computational cost of MDT could be significantly larger than the cost of CMI, given the availability of advanced computational resources, the computational efficiency  can be traded for more reliable results. In our simulated experiments, the running time of CMI was in order of seconds while the running time of MDT was in order of minutes, using a 2.6 GHz quad core CPU and 16 GB RAM. 

Finally, we discuss the assumptions behind our method, thus outlining situations where the method is expected to perform well. 
There are four key ingredients in our method:
\begin{enumerate}
 \item We perform independence tests on time series pairs. There is no guarantee that if there were higher-order dependencies among several time-series (e.g. the outcome variable and two confounding variables), they would be detected by the pair-wise tests. 
 \item In our conditional independence testing, the maximum conditioning set is determined by the largest lag $L$. The value of $L$ is of course upper bounded by our desire to have sample sizes large enough to yield sufficient power to independence tests. 
 \item The matching procedure assumes that there is an overlap in the confounding variables' values between the groups of treated and control units. If this is not the case, the matching will not achieve sufficient balance. For example, in our case, it was not feasible to conduct a causality study of the impact of social media sentiment on the treated assets prices of Apple Inc. when the $T3$ thresholds pair is used, since there was not sufficient overlap on the confounding variables' values. 
 \item The estimation of the average treatment effect is influenced by the power of the statistical test that is applied.     
\end{enumerate}

\section{Conclusions}
\label{sec:conclusions}

In this study, for the first time we have attempted to quantify the causal impact of social and emotional factors, captured by social media, on daily stock market returns of individual companies (i.e., not just a mere correlation between the two). We have proposed a novel non-parametric framework for causal analysis in time-series. Our evaluation on synthetic data demonstrates that our method is more effective on inferring true causality and avoiding false positive conclusions compared to other methods that have been previously used for causal inference in time-series. Our approach can incorporate a large number of factors and, therefore, can effectively handle complex data such as financial data. Indeed, causality studies that are based on observational data rather than experimental procedures could be biased in case of missing confounding variables. However, conducting experimental studies is not feasible in most cases. In this work we have minimized the risk of biased conclusions due to unmeasured confounding variables by including in our study a large number of factors. Additionally, we conduct an analysis on the sensitivity of our conclusions on missing confounding variables. 
We have estimated a sentiment index indicating the probability that the general sentiment of a day, based on tweets posted for a target company, is positive. Our results show that Twitter data polarity does indeed have a causal impact on the stock market prices of the examined companies. 
Although our study involves only big technological companies and consequently our findings might not be generalizable, especially for companies that do not focus directly on retail customers. The analysis of the validity of these findings in the B2B sector is an open question that we plan to explore. Hence, we believe social media data could represent a valuable source of information for understanding the dynamics of stock market movements. 

% \bibliography{myBib}
% \bibliographystyle{unsrt}

% That's all folks!
\end{document}